\documentclass{article}

    \usepackage[preprint]{neurips_2025}

\usepackage[utf8]{inputenc} 
\usepackage[T1]{fontenc}    
\usepackage{hyperref}       
\usepackage{url}            
\usepackage{booktabs}       
\usepackage{amsfonts}       
\usepackage{nicefrac}       
\usepackage{microtype}      
\usepackage{xcolor}         

\usepackage{amsmath}
\usepackage{mathtools}
\usepackage{amssymb}
\usepackage{amsthm}
\usepackage[linesnumbered,ruled,vlined]{algorithm2e}
\usepackage{bbm} 
\usepackage{tabularx}
\usepackage{longtable}
\usepackage{booktabs}

\DeclareSymbolFont{bbold}{U}{bbold}{m}{n}
\DeclareSymbolFontAlphabet{\mathbbold}{bbold}

\newtheorem{theorem}{Theorem}[section]

\newtheorem{proposition}[theorem]{Proposition}
\newtheorem{lemma}[theorem]{Lemma}
\newtheorem{example}{Example}[section]
\newtheorem{definition}[theorem]{Definition}
\newtheorem{assumption}[theorem]{Assumption}

\usepackage{listings} 
\lstdefinestyle{pythonstyle}{
  language=Python,
  basicstyle=\ttfamily\small,
  keywordstyle=\color{blue},
  commentstyle=\color{gray},
  stringstyle=\color{red},
  numbers=left,
  numberstyle=\tiny\color{gray},
  stepnumber=1,
  numbersep=5pt,
  showspaces=false,
  showstringspaces=false,
  showtabs=false,
  frame=single,
  tabsize=2,
  breaklines=true,
  breakatwhitespace=false,
  escapeinside={\%*}{*)}
} 

\lstdefinestyle{terminalstyle}{
  basicstyle=\ttfamily\small,    
  frame=single,                  
  breaklines=true,               
}

\title{Information Bargaining: Bilateral Commitment in Bayesian Persuasion}

\author{%
  Yue Lin \\
  The Chinese University of Hong Kong, Shenzhen \\
  \texttt{yuelin@link.cuhk.edu.cn} \\
  \And
  Shuhui Zhu \\
  University of Waterloo \\
  \texttt{shuhui.zhu@uwaterloo.ca} \\
  \And
  William A Cunningham \\
  University of Toronto \\
  Vector Institute \\
  \texttt{wilcunningham@gmail.com} \\
  \And
  Wenhao Li \\
  Tongji University \\
  \texttt{whli@tongji.edu.cn} \\
  \And
  Pascal Poupart \\
  University of Waterloo \\
  \texttt{ppoupart@uwaterloo.ca} \\
  \And
  Hongyuan Zha \\
  The Chinese University of Hong Kong, Shenzhen \\
  \texttt{zhahy@cuhk.edu.cn} \\
  \And
  Baoxiang Wang \\
  The Chinese University of Hong Kong, Shenzhen \\
  \texttt{bxiangwang@cuhk.edu.cn}
}

\begin{document}

\maketitle

\begin{abstract}
Bayesian persuasion, an extension of cheap-talk communication, involves an informed sender committing to a signaling scheme to influence a receiver’s actions. Compared to cheap talk, this sender's commitment enables the receiver to verify the incentive compatibility of signals beforehand, facilitating cooperation. 
While effective in one-shot scenarios, Bayesian persuasion faces computational complexity (NP-hardness) when extended to long-term interactions, where the receiver may adopt dynamic strategies conditional on past outcomes and future expectations.
To address this complexity, we introduce the bargaining perspective, which allows:
(1) a unified framework and well-structured solution concept for long-term persuasion, with desirable properties such as fairness and Pareto efficiency;
(2) a clear distinction between two previously conflated advantages: the sender’s informational advantage and first-proposer advantage.
With only modest modifications to the standard setting, this perspective makes explicit the common knowledge of the game structure and grants the receiver comparable commitment capabilities, thereby reinterpreting classic one-sided persuasion as a balanced information bargaining framework.
The framework is validated through a two-stage validation-and-inference paradigm: We first demonstrate that \texttt{GPT-o3} and \texttt{DeepSeek-R1}, out of publicly available LLMs, reliably handle standard tasks; We then apply them to persuasion scenarios to test that the outcomes align with what our information-bargaining framework suggests.
All code, results, and terminal logs are publicly available at \url{https://github.com/YueLin301/InformationBargaining}.
\end{abstract}


\section{Introduction}

Communication plays a pivotal role in human society, shaping interactions and influencing decision-making processes. The concept of ``cheap talk'', introduced by~\citet{crawford1982strategic}, underscores the importance of verbal communication in strategic contexts. 
McCloskey's assertion that ``one quarter of GDP is persuasion'' illustrates how integral persuasion is to economic activities~\citep{mccloskey1995one}. Recent analyses by~\citet{antioch2013persuasion} suggest that this figure has risen to $30\%$, highlighting the growing significance of persuasive communication in the modern economy. 
In this landscape, the idea of Bayesian persuasion, proposed by~\citet{kamenica2011bayesian}, emerges as a crucial framework, involving the transmission of signals to influence a receiver's beliefs and actions. This framework finds a wide range of applications across various domains, including routing systems~\citep{das2017reducing, kremer2014implementing}, law enforcement deployment~\citep{hernandez2022bayesian, lazear2006speeding}, and grading in schools~\citep{boleslavsky2015grading, ostrovsky2010information}, as surveyed by~\citet{kamenica2019bayesian}.

A series of studies through the lens of Bayesian Correlated Equilibrium (BCE) demonstrates the high relevance of Bayesian persuasion. Following \citet{bergemann2013robust,bergemann2016bayes}, Bayesian persuasion can be understood as the design of an optimal Bayes-correlated equilibrium: the sender \textbf{commits} \textit{ex ante} to a signaling scheme that the receiver will find incentive-compatible to follow \citep{bergemann2019information}. As noted by~\citet{dughmi2017algorithmic}, the commitment assumption is essential to Bayesian persuasion, otherwise the model will degenerate to the cheap talk model~\citep{crawford1982strategic}. Compared to the analysis of partition equilibrium in cheap talk, which suggests that cooperation becomes impossible when the conflict of interest is too large, the introduction of commitment enables concavification~\citep{kamenica2011bayesian} and Bayes Correlated Equilibrium (BCE) analysis to support deeper cooperation in a broader range of scenarios.

\begin{figure}[thbp]
    \centering
    \includegraphics[width=\textwidth]{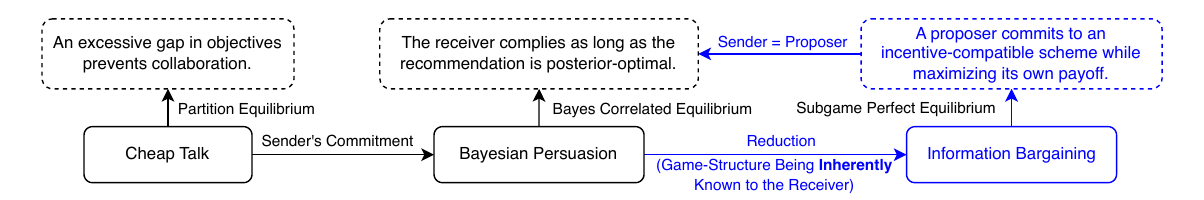}
    \caption{The framework of communication with charge-free signals and information asymmetry. The solid-bordered boxes represent the games, while the dashed-bordered boxes represent the corresponding analyses. The parts highlighted in \textcolor{blue}{blue} are our contributions.}
    \label{fig: Framework}
\end{figure}

While effective in one-shot scenarios, Bayesian persuasion faces significant complexity (NP-hardness) in long-term interactions, where receivers may adopt dynamic strategies conditional on past outcomes and future expectations~\citep{gan2022bayesian}, prompting various online strategic proposals. For example, \citet{gan2022bayesian} developed threat-based meta signaling schemes, and \citet{bernasconi2024persuading} explored promise-form signaling schemes, prompting various strategic proposals. For example, \citet{gan2022bayesian} developed threat-based meta signaling schemes, and \citet{bernasconi2024persuading} explored promise-form signaling schemes.
Despite these advances, the literature currently lacks a unified theoretical framework capable of systematically integrating these diverse strategies. 

To address this gap, we introduce the bargaining perspective to Bayesian persuasion. This perspective reframes classic one-sided persuasion as a balanced information bargaining framework by explicitly acknowledging the common knowledge of the game structure and providing the receiver comparable commitment capabilities. With only modest modifications to standard assumptions, our bargaining framework achieves two primary goals:
\textbf{(1)} It provides a unified framework and well-structured solution concepts for analyzing long-term persuasion, characterized by desirable properties such as fairness and Pareto efficiency~\citep{nash1950bargaining}.
\textbf{(2)} It clarifies two previously conflated advantages inherent to Bayesian persuasion: the sender’s informational advantage and the first-proposer advantage.
These let us isolate a bargaining stage without changing optimality or equilibrium in long-term Bayesian persuasion, so the players’ interaction resembles signing a \textit{contract}. When Bayesian persuasion is used in practice, we expect receivers to bargain if it knows the sender’s reward function and if the sender lacks a commitment device that secures the first-proposer advantage. Viewed this way, the existing online persuasion strategies can be interpreted as bargaining tactics.

To empirically evaluate our information bargaining framework, we leverage recent advances in LLMs, which have demonstrated substantial improvements in reasoning capabilities~\citep{guo2025deepseek}.
These models are increasingly employed as solvers for complex strategic problems. In our study, we use LLMs as equilibrium solvers to directly test this framework. 
We begin by evaluating several LLMs on tasks with known theoretical results to assess their reasoning capabilities. Among them, \texttt{GPT-o3}\citep{openai2025o3} and \texttt{DeepSeek-R1}\citep{guo2025deepseek} pass our validation criteria. We subsequently apply these validated models to compute equilibria within our bargaining framework, with results supporting our claims.

\section{Preliminaries}

To be self-contained, we briefly introduce the frameworks of cheap talk, Bayesian persuasion, and bargaining in this section. All notations are summarized in the glossary (Appendix~\ref{Section: Glossary}), and the full procedures of all games are included in Appendix~\ref{Section: Game Procedures} to facilitate cross-game comparison.

\subsection{Cheap Talk and Bayesian Persuasion}
\label{persuasion preliminaries}

Cheap talk models cost-free, non-binding messages between two players~\citep{crawford1982strategic,farrell1996cheap}. Procedure \ref{Algorithm: Cheap Talk} in Appendix \ref{Section: Game Procedures} (identical to Procedure \ref{Algorithm: Bayesian Persuasion} but without the blue commitment steps) formalizes the game. In perfect Bayesian Nash equilibrium, such talk can sustain cooperation when goals are not too misaligned~\citep{crawford1982strategic,crawford1998survey}.

Bayesian persuasion introduces a \textbf{commitment assumption}: the sender commits to and then follows a signaling scheme chosen before the state is realized~\citep{kamenica2011bayesian,kamenica2019bayesian}. Commitment may arise from repeated-game reputation~\citep{dughmi2016algorithmic,dughmi2017algorithmic}, institutional authority (e.g., grading policies~\citep{dughmi2016algorithmic} or legislation~\citep{kamenica2019bayesian}). This additional lever enlarges the set of cooperative equilibria relative to cheap talk~\citep{kamenica2011bayesian}.
The procedure of Bayesian persuasion is described as Procedure~\ref{Algorithm: Bayesian Persuasion}, where a sender $i$ is trying to send signals $\sigma \in \mathbbold{\Sigma}$ to influence the actions $a\in \mathbb{A}$ of a receiver $j$. 
(1) A neutral environment will first samples a state $s \in \mathbb{S}$ according to a prior distribution $\mu_0 \in \Delta(\mathbb{S})$. State $s$ is observable only for the sender, and it influence both players' payoffs. So the receiver has interests in state $s$ without knowing it, which gives an opportunity to the sender to influence the behavior of the receiver. This refers to the sender's \textbf{informational advantage.}
(2) Next, the sender sends a signal $\sigma \in \mathbbold{\Sigma}$ to the receiver. Its signaling scheme is defined as $\varphi: \mathbb{S} \to \Delta(\mathbbold{\Sigma})$, and the signaling scheme set is $\mathbbold{\Psi}$.
(3) Then, the receiver takes an action $a \in \mathbb{A}$ given the signal. Its action rule is defined as $\pi: \mathbbold{\Sigma} \to \mathbb{A}$, and the action rule set is $\mathbbold{\Pi}$.
The receiver's behavior is default to be Bayesian, which gives the task name ``Bayesian persuasion''. 
In the cheap talk setting, the receiver behaves in the same way: after receiving a signal, it updates its posterior belief and chooses the optimal strategy based on that belief.
(4) The payoff functions of the sender and receiver are defined as $r^i: \mathbb{S} \times \mathbb{A} \to \mathbb{R}$ and $r^j: \mathbb{S} \times \mathbb{A} \to \mathbb{R}$, respectively. They do not generally share the same payoff function, so the task is mixed-motive at the most time. 
Any payoff instance is independent to the sender's strategy value $\sigma$, so for the sender to get better payoffs, it must leverage its informational advantage to manipulate the receiver to act towards its (the sender's) favor.

\begingroup
\renewcommand{\algorithmcfname}{Procedure}
\begin{algorithm}
\label{Algorithm: Bayesian Persuasion}
\caption{Timing of Bayesian Persuasion}
\KwIn{Game settings $(\mu_0, \mathbb{S}, \mathbbold{\Sigma}, {\mathbb{A}}, r^i, r^j)$}

{\fontsize{10.15pt}{\baselineskip}\selectfont

\textcolor{blue}{The sender decides a signaling scheme $\varphi$}
\tcp*{Commitment (1)}

\textcolor{blue}{The sender commits $\varphi$ to the receiver}
\tcp*{Commitment (2)}

Environment samples a state $s \sim \mu_0(\cdot)$ \;

The sender signals $\sigma \sim \varphi(\cdot \mid s)$ \textcolor{blue}{as committed}
\tcp*{Commitment (3)}

The receiver takes an action $a \sim \pi(\cdot \mid \sigma)$ \;

The sender and the receiver get rewards of $r^i(s, a)$ and $r^j(s, a)$ respectively \;
}
\end{algorithm}
\endgroup


As stated, the receiver needs to make a Bayesian best response, which requires knowing the sender's signaling scheme $\varphi$. This is allowed in Bayesian persuasion and is referred to as the commitment assumption. 
In cheap talk, the receiver does not know the sender’s signaling scheme in advance; it is only used in the analysis of equilibrium.

\begin{assumption}[Commitment]
\label{Assumption: Commitment Assumption}
The sender commits to a signaling scheme $\varphi$, which encompasses the following three key aspects: (1) The sender will decide on a $\varphi$ before the game starts, (2) The sender will honestly inform the receiver of this $\varphi$, and (3) during the game, the sender will actually samples signals $\sigma \in \mathbbold{\Sigma}$ according to the committed $\varphi$.
\end{assumption}


We can further simplify the problem without loss of generality by assuming that the sender ``sending a signal'' is equivalent to ``recommending an action'' for the receiver to take. And this is known as an analysis similar to the revelation principle, proposed and proved in~\citet{kamenica2011bayesian}. 
\begin{proposition}[A Variant of Revelation Principle, Paraphrased from \citet{dughmi2017algorithmic}]
Assuming that the signaling set is equal to the action set ($\mathbbold{\Sigma} = \mathbb{A}$) does not affect the optimality of the sender's signaling scheme $\varphi$. This implies that every signaling scheme is equivalent to one that recommends actions.
\footnote{We will refer to this as the revelation principle in the following content since there will be no ambiguity here. We keep using the notation $\sigma$ to denote an action sent by the sender, using $a$ to denote an action actually taken by the receiver.}
\label{Proposition: Revelation Principle}
\end{proposition}


Let $\mathbb{V}^i$ and $\mathbb{V}^j$ denote the visibility set of the sender and receiver, respectively. Let $\mathbb{V}^{i+j}$ denote the common knowledge set (known by the players, with each aware that others know it, ad infinitum). Then $\mathbb{V}^i \setminus \mathbb{V}^j = \{s\}$ and $i, j, \mu_0, \mathbb{S}, \mathbb{A}, r^j, \varphi, \sigma, a  \in \mathbb{V}^{i+j}$. The visibility of $r^i$ is not specified in previous work. Here we assume $r^i \in \mathbb{V}^{i+j}$.



To further characterize the task, we can restrict our attention without loss of generality to a subset of $\mathbbold{\Phi}$ whose elements are \textit{signaling schemes that the receiver would respect}.
This can be clearly introduced by the concept of \textbf{incentive compatibility}.


\begin{definition}[Incentive Compatibility in Bayesian Persuasion, Single-Receiver Version without Types of~\citet{bergemann2016bayes, dughmi2017algorithmic}]
\label{Definition: Incentive Compatibiliy in Bayesian Persuasion}
The sender's recommendation $\phi$ is incentive compatible for the receiver if for each $a \in \mathbb{A}$,
\begin{equation}
\sum\limits_{s} \mu_0(s) \cdot \varphi(\sigma=\sigma \mid s) \cdot r^j(s, a) 
\geq \sum\limits_{s} \mu_0(s) \cdot \varphi(\sigma \mid s) \cdot r^j(s, a'),
\end{equation}
for all $a' \in \mathbb{A}$.
\end{definition}

Every signal (recommended action) will induce a posterior belief of the receiver, who will then choose the Bayesian best response to it. If the receiver’s best response is exactly the action recommended by the sender, then the receiver is considered \textit{persuaded} by the sender. 
In this case, the receiver has no incentives to deviate from the sender's recommendation, i.e., $\pi(a\mid\sigma=a)=1$, and thus can be omitted in this canonical setting. In the origin paper~\citet{bergemann2016bayes}, it is named as the ``obedience''.
\textbf{Bayes correlated equilibria (BCE)} is defined as the set of obedient signaling schemes.


Having these, the sender faces an optimization problem in the beginning of the game (Line 1 in Procedure~\ref{Algorithm: Bayesian Persuasion}):
\begin{equation}
\label{Equation: Canonical BP Linear Program}
\begin{aligned}
\max_{\varphi} \quad & \mathbb{E}_{\mu_0, \varphi, \pi}\left[r^i(s, a)\right] := \sum_{s} \mu_0(s) \sum_{a} \varphi(\sigma=a\mid s) \cdot r^i(s, a) \\
\text{s.t.} \quad & \sum_{s} \mu_0(s) \cdot \varphi(\sigma=a \mid s) \cdot \left[ r^j(s, a) - r^j(s, a')\right] \geq 0, \quad \forall a, a' \in \mathbb{A}.
\end{aligned}
\end{equation}
That is, the sender is to select an incentive compatible signaling scheme for the receiver, to maximize its (the sender's) own expected payoffs.
We refer to the canonical Bayesian persuasion as the formulation in~\citet{dughmi2017algorithmic}. Concrete examples of persuasion are provided in Section~\ref{Tasks of Persuasion and Bargaining}.

\begin{definition}[Bayesian Persuasion]
\label{Definition: Bayesian Persuasion}
Under Assumption~\ref{Assumption: Commitment Assumption} and Proposition~\ref{Proposition: Revelation Principle}, a Bayesian persuasion task is defined as $\mathcal{BP}:=(\mathbb{S}, \mu_0, \mathbb{A}, r^i, r^j)$, where a sender tries to persuade a receiver according to Procedure~\ref{Algorithm: Bayesian Persuasion}.
\end{definition}

\subsection{Bargaining Game}


The bargaining game, initially introduced by \citet{nash1950bargaining, nash1953two}, serves as a foundational model in the study of strategic negotiation. We first introduce the definition in~\citet{Maschler_Solan_Zamir_2013}.
\begin{definition}[Cooperative Bargaining Game~\citep{Maschler_Solan_Zamir_2013}]
\label{Definition: Bargaining}
A two-player bargaining game is defined as $\mathcal{BG} := (\mathbb{Y}, \boldsymbol{d})$, where $\mathbb{Y} \subseteq \mathbb{R}^2$ is a nonempty, compact \textbf{feasibility set} of possible agreements, and $\boldsymbol{d} = (d_i, d_j) \in \mathbb{R}^2$ is the \textbf{disagreement point}. It is assumed that some agreement $y = (y_i, y_j) \in \mathbb{Y}$ strictly improves upon disagreement for both players, i.e., $y_i > d_i$ and $y_j > d_j$.
\end{definition}

A critical element of the bargaining game is the presence of a disagreement point, which fundamentally enables each player to unilaterally refuse cooperation. 
Concrete examples of bargaining are provided in Section~\ref{Tasks of Persuasion and Bargaining}.

This definition does not specify the details of game dynamics and is typically referred to as cooperative bargaining. The solution concept in this setting is often given by axiomatic models, which propose certain axioms as necessary and sufficient conditions for a solution to the optimization problem. Notable examples include the Nash bargaining solution~\citep{nash1950bargaining}, the Kalai–Smorodinsky bargaining solution~\citep{kalai1975other}, and the proportional bargaining solution~\citep{kalai1977proportional}. These axioms include several desirable properties, such as fairness and Pareto efficiency~\citep{nash1950bargaining}.

\begin{definition}[Solution Concept of Cooperative Bargaining Game~\citep{Maschler_Solan_Zamir_2013}]
\label{Definition: Bargaining Solution}
A solution concept of bargaining games is defined as a function $\omega_{\mathcal{BG}}
$ which maps
every bargaining game $(\mathbb{Y}, \boldsymbol{d})$
to an agreement $\boldsymbol{y} \in \mathbb{Y}$.
\end{definition}

Another class of bargaining games with clearly defined dynamics, particularly suitable for long-term interactions, is \textbf{Rubinstein’s alternating-offer bargaining model}~\citep{rubinstein1982perfect}. The game is an extensive-form game, and its procedure is described in Algorithm~\ref{Algorithm: Rubinstein Bargaining} in Appendix~\ref{Section: Procedure of Rubinstein's Alternating-Offer Bargaining Game}. In this model, two players participate, and the proposer is initially selected at uniformly random. The proposer suggests how to divide a certain amount of surplus. If the responder accepts, both players receive the proposed payoffs. If not, they switch roles, and the original responder becomes the proposer in the next round. 
When the number of stages in this model is one at most, then the model is referred to as a \textbf{Stackelberg bargaining game} or an \textbf{ultimatum game}.
This type of game has well-defined solution concepts, i.e. subgame perfect Nash equilibrium. By introducing a patience factor, one can derive explicit expected payoffs for each player and, under certain parameter values, obtain results that align with those from axiomatic models. As a result, the model provides both a rigorous solution and desirable beneficial properties.

\section{Revisiting Bayesian Persuasion}

In this section, we revisit the canonical setting of one-shot Bayesian persuasion without introducing any new mechanisms. 
Essentially we are focusing on the Procedure~\ref{Algorithm: Bayesian Persuasion}.
The idea is sketched as follows. 
First we argue that the receiver has sufficient knowledge of the game structure that it can calculate its expected payoff once the sender commits to a signaling scheme. 
Then we show that there always exists a disagreement point in Bayesian persuasion where the receiver can ignore the sender.
At the disagreement point, the receiver can unilaterally set both players' payoffs to default values, which nullifies the sender's informational advantage.
From this intuition, Bayesian persuasion can be roughly interpreted as an ultimatum game.

\subsection{The Receiver's Awareness of the Game Structure}

We start with recalling how the receiver will act before they reach a consensus,
i.e., $\pi(a \mid \sigma=a)$ does not have to be $1$ and  thus cannot be simplified.

Since
$\mu_0, \varphi \in \mathbb{V}^{i+j}$,
given a $\pi$,
the Bayesian receiver is able to calculate its ground truth expected payoff as 
$
\sum_{s}\mu_0(s) 
\sum_{\sigma} \varphi(\sigma\mid s) 
\sum_{s'} \mu(s' \mid \sigma) 
\sum_{a} \pi(a\mid s') \cdot 
r^j(s, a)
$,
where 
$
\mu(s\mid \sigma) = 
{\left(\mu_0(s)\cdot \varphi(\sigma\mid s)\right)}/{\left(\sum_{s'} \mu_0(s') \cdot \varphi(\sigma\mid s')\right)}
$
is the posterior belief according to Bayes' rule, and $\pi$ optimizes the receiver's payoff from its posterior view. 
Or more generally for an arbitrary decision rule $\pi$, the receiver's ground true expected payoff is
\begin{equation}
\label{Equation: Receiver's Expected Payoff}
R^j(\mu_0, \varphi, \pi, r^j)
:= \sum\limits_{s}\mu_0(s) 
\sum\limits_{\sigma} \varphi(\sigma\mid s) 
\sum\limits_{a} \pi(a\mid \sigma) \cdot 
r^j(s, a).
\end{equation}
Because the receiver can see the sender's reward function ($r^i \in \mathbb{V}^{i+j}$), the receiver can also calculate the sender's expected payoff, by replacing $r^j(s, a)$ in Equation~\eqref{Equation: Receiver's Expected Payoff} with $r^i(s, a)$.

Define the \textbf{game structure} for every task $\mathcal{BP}$ as a mapping $\psi: \mathbbold{\Phi}\times \mathbbold{\Pi} \to \{\boldsymbol{R}\}$, where $\{\boldsymbol{R}\}:= \{ (R^i, R^j) \}$ is the set of all possible expected payoff pairs in $\mathcal{BP}$. 
The game structure is determined by the nature of the parameters and the intrinsic setting of Bayesian persuasion. It therefore can be reasoned out if the task parameters are specified.
We can now conclude that $\psi \in \mathbb{V}^{i+j}$, since it can be reasoned out by both players using other common knowledge. 
Let $\mathbbold{\Psi}$ denote all the possible game structure mappings for different $\mathcal{BP}$ tasks.

\subsection{Action Rule with Inherent Game Structure Awareness}
\label{Section: Action Rule with Game Structure Awareness}

The receiver's awareness of the game structure is important, as it allows more complicated action rules.
We define the \textbf{action rule with game structure awareness} as $\tilde{\pi}: \mathbbold{\Psi} \times \mathbbold{\Phi} \times \mathbbold{\Sigma} \to \Delta(\mathbb{A})$, which enables Receiver to decide its $\pi$ depending on both the game structure $\psi$ and the committed signaling scheme $\varphi$.

Recall that, given a committed signaling scheme $\varphi$ and a signal $\sigma$, the receiver will first calculate its posterior belief and then choose its best response to this. We call this default Bayesian best response as the action rule $\pi_1:=\mathrm{BestResponseTo}(\mathcal{BP}, \varphi, \sigma)$. The corresponding expected payoff pairs is denoted as $\boldsymbol{R}_1 = (R^i_1, R^j_1) = \psi(\varphi, \pi_1)$.

In all cases, the receiver always has the choice to ignore the sender's signaling scheme and signals. It can calculate its best response solely on its prior knowledge. We denote the prior best response as the decision rule $\tilde{\pi}_0:=\mathrm{BestResponseTo}(\mathcal{BP})$. The corresponding expected  payoff pairs is denoted as $\boldsymbol{R}_0 = (R^i_0, R^j_0) = \psi(\cdot, \pi_0)$, since it is independent of $\varphi$. It means that the receiver has the ability to \textbf{unilaterally} determine both players' expected payoffs, which aligns with the intuition of the disagreement points in bargaining games. This is also referred to the babbling equilibrium in the cheap talk model.

Having these, we now consider an example $\tilde{\pi}$: Satisfaction Check, as described in Algorithm~\ref{Algorithm: Satisfaction Check}.
The \textbf{satisfaction threshold function} is defined as an indicator function $\chi: \{\boldsymbol{R}\}^2 \to \{0, 1\}$ and reflects the receiver's non-myopic preferences, such as reflecting the receiver's demands for fairness. 
Then if the receiver is satisfied with the payoffs ($\chi(\boldsymbol{R}_0, \boldsymbol{R}_1)=1$), then it chooses $\tilde{\pi}(\cdot \mid \psi, \varphi, \sigma)=\pi_1(\cdot \mid \sigma)$ otherwise $\tilde{\pi}(\cdot \mid \psi, \varphi, \sigma)=\pi_0(\cdot \mid \sigma)$. Some more specific examples of $\chi$ are provided below Algorithm~\ref{Algorithm: Satisfaction Check} in Appendix~\ref{Section: Procedure of Satisfaction Check}.




\section{Information Bargaining}

In long-term settings, we show that Bayesian persuasion divides into two sequential stages: (1) a bargaining phase where the players negotiate a signaling scheme and (2) a realization phase where they act on the agreed signals. This perspective separates the sender’s informational advantage from the first-mover advantage, with the latter providing most of the leverage. Because first-mover privileges do not arise automatically in repeated interactions, practical applications must introduce explicit commitment devices; otherwise, persuasion loses much of its force. Recognizing this need, we recast Bayesian persuasion as a cooperative bargaining game and motivate a solution concept that also anticipates possible retaliation by the receiver.




\subsection{Long-Term Bayesian Persuasion}
\label{Section: Long-Term Bayesian Persuasion}

Without changing optimality or equilibrium, long-term Bayesian persuasion can be decomposed into two distinct stages: a bargaining stage and a realization stage. The detailed development and key insights are presented in Appendix~\ref{Section: Procedures of Long-Term Bayesian Persuasion}; here, we briefly sketch the main idea. We have discussed how the receiver can adopt a more sophisticated action rule to express its attitude toward the sender. For instance, the receiver might temporarily refuse to cooperate in the short term to pressure the sender into being more honest in the long term. This insight is akin to the ``tit-for-tat'' strategy in the repeated prisoner's dilemma~\citep{axelrod1981evolution}. Moreover, since both parties are aware of the game's structure, once they have established a signaling scheme $\varphi$ and an action rule $\pi$, they can compute their expected payoffs without needing to sample or gather feedback. Therefore, the current study of long-term Bayesian persuasion can be interpreted in a naive repeated style and can be described as Procedure~\ref{Algorithm: Long-Term Bayesian Persuasion (3)}.

\begingroup
\renewcommand{\algorithmcfname}{Procedure}
\begin{algorithm}
\label{Algorithm: Long-Term Bayesian Persuasion (3)}
\caption{Timing of Long-Term Bayesian Persuasion}
\KwIn{Game settings $(\mu_0, \mathbb{S}, \mathbbold{\Sigma}, {\mathbb{A}}, r^i, r^j)$, number of interactions $T_b$, $T_r$}

{\fontsize{10.15pt}{\baselineskip}\selectfont


\For{$t = 1$ \textbf{to} $T_b$}{
    The sender decides and \textcolor{blue}{declares} a signaling scheme $\varphi$ \;

    The receiver decides and \textcolor{blue}{declares} an action policy $\pi$ given $\varphi$ \;
}
\For{$t_s = 1$ \textbf{to} $T_r$}{
    Environment samples a state $s \sim \mu_0(\cdot)$ \;

    The sender signals $\sigma \sim \varphi(\cdot \mid s)$ \textcolor{blue}{according to the final decision} \;

    The receiver takes an action $a \sim \pi(\cdot \mid \sigma)$ \textcolor{blue}{according to the final decision} \;

    The sender and the receiver get rewards of $r^i(s, a)$ and $r^j(s, a)$ respectively \;
}
}
\end{algorithm}
\endgroup

From this, we can see that from the perspective of long-term interaction, the receiver can also commit to an action rule before the realization stage just like the sender, since the actual game takes place during the bargaining stage while the realization stage merely executes the outcome to obtain payoffs. So the commitment assumption that distinguishes Bayesian persuasion from cheap talk applies not only to the sender but also to the receiver.

\subsection{Two Conflated Sender Advantages}

Now we discuss two advantages of the sender that have been conflated in previous studies: the informational advantage and the first-proposer advantage. As we can see from Procedure~\ref{Algorithm: Long-Term Bayesian Persuasion (3)}, the sender commits first. Therefore, from the perspective of analyzing subgame perfect equilibrium using backward induction, a rational receiver will always accept. As a result, even if the bargaining stage involves repeated interactions, the outcome is essentially the same as in a one-shot interaction that of canonical Bayesian persuasion.
This outcome is due to the sender committing first. From a rationality-based perspective, there is not much room to explore in this setting; most existing studies instead focus on behavioral strategies.
A more detailed review of related work that embeds bargaining strategies within Bayesian persuasion is provided in Appendix~\ref{Evidence of Bayesian Persuasion Being a Bargaining Game}.

In the bargaining stage, the receiver can commit first as well, since this is independent of the realization order of the signal and the action. In this way, the sender must consider whether to fulfill the receiver's expectations. This is recognized as \textit{the power of commitment}~\citep{nowak2000fairness}. 
Detailed discussions can be found in Appendix~\ref{Section: Receiver Commitment}.

\subsection{Joint Commitment and the Solution Concept of Bayesian Persuasion}

Now we define the solution concept of Bayesian persuasion by introducing the idea of joint commitment.
To eliminate the players’ first-proposer advantage, we consider modifying Procedure 2 such that the bargaining stage becomes a simultaneous-move game, where both players declare their policies at the same time. In the next round, they update their policies based on the declared profile from the previous round. The strategy-updating process of each player can be described by a function $f: \mathbbold{\Phi} \times \mathbbold{\Pi} \to \mathbbold{\Phi} \times \mathbbold{\Pi}$, and then the outcome of the bargaining stage can then be interpreted as a fixed point of the dynamic function $f$. We define the fixed point as the joint commitment.

\begin{definition}[Joint Commitment]
\label{Definition: Joint Commitment}
A joint commitment of a Bayesian persuasion task $\mathcal{BP}= (\mathbb{S}, \mu_0, \mathbb{A}, r^i, r^j)$ is a fixed point of a dynamic function $f: \mathbbold{\Phi} \times \mathbbold{\Pi} \to \mathbbold{\Phi} \times \mathbbold{\Pi}$, where $f$ represents the strategy-updating process of players in the simultaneous-move bargaining stage, 
$\mathbbold{\Phi} := \{\varphi: \mathbb{S} \to \Delta(\mathbbold{\Sigma})\}$ ,
and
$\mathbbold{\Pi} := \{\pi: \mathbbold{\Sigma} \to \Delta(\mathbb{A})\}$.
\end{definition}


We then extend the conclusion from Section~\ref{Section: Action Rule with Game Structure Awareness}: not only can the receiver unilaterally enforce the babbling equilibrium, but the sender can as well, by specifically choosing $\varphi$ such that $\mu = \mu_0$ (for instance, regardless of the current state $s$, the sender consistently sends a dummy signal $\sigma_0$, i.e., $\varphi(\sigma_0 \mid s) = 1, \forall s$).
In this case, $\pi_1 = \pi_0$ and thus $\boldsymbol{R}_1 = \boldsymbol{R}_0$. We denote this type of signaling scheme as $\varphi_0$.
Having these in place, we define the solution concept of Bayesian persuasion as follows.

\begin{definition}[Solution Concept of Bayesian Persuasion]
\label{Definition: Persuasion Solution}
A solution concept of Bayesian persuasion $\mathcal{BP} = (\mathbb{S}, \mu_0, \mathbb{A}, r^i, r^j)$ is a function $\omega_{\mathcal{BP}}$
which maps every
$\mathcal{BP}$
to a joint commitment
$(\varphi, \pi) \in \mathbbold{\Phi} \times \mathbbold{\Pi}$, where 
$\varphi^* \ne \varphi_0$, 
$\pi^* \ne \pi_0$,
$\mathbbold{\Phi} := \{\varphi: \mathbb{S} \to \Delta(\mathbbold{\Sigma})\}$ ,
and
$\mathbbold{\Pi} := \{\pi: \mathbbold{\Sigma} \to \Delta(\mathbb{A})\}$.
\end{definition}

Then it is straightforward to prove the following reduction: Bayesian persuasion is reducible to a cooperative bargaining game. The detailed proof is provided in Appendix~\ref{Section: Proof of Reduction from Bayesian Persuasion to Bargaining Games}.

\begin{lemma}[Reduction from Bayesian Persuasion to Bargaining Games]
\label{Lemma: BP_Is_a_BG}
When the Assumption~\ref{Assumption: Existence of Better Outcomes} holds, a Bayesian persuasion task $\mathcal{BP}=(\mathbb{S}, \mu_0, \mathbb{A}, r^i, r^j)$ is polynomial-time reducible to a bargaining game $\mathcal{BG}=(\mathbb{Y}, \boldsymbol{d})$. That is, $\mathcal{BP} \leq_p \mathcal{BG}$.
\end{lemma}

We provide in Appendix~\ref{Section: Advantages of Information Bargaining Framework} an example of how the Nash bargaining solution can be used to solve Bayesian persuasion: the optimization problem~\ref{Equation: Canonical BP Linear Program} can be reformulated as maximizing the Nash product. We also include a discussion of some advantages brought by this framework.

\section{Empirical Results}

In this section, we aim to demonstrate 
\textbf{(a)} the receiver is aware of the game structure and is capable of constructing a more sophisticated action rule;
\textbf{(b)} the sender’s behavioral dominance arises not only from the sender possessing private information, but also from the sender’s role as the first proposer. 
We validated these two hypotheses using LLMs across 87 different settings. 
Among them, 62 settings have clear theoretical analyses and were used to evaluate the reasoning capabilities of the LLMs to ensure that the selected models meet the required standards. Only after this validation step did we use verified LLMs to test the hypotheses. We evaluated 5 LLMs in total, of which 2 passed the test: their outputs showed sufficiently high correlation with peer-reviewed theoretical results. These models are all reasoning models, which are \texttt{GPT o3-2025-04-16} and \texttt{deepseek-reasoner} (R1 version).
Using these validated models, the experiments show that the results across various long-term persuasion settings are consistent with those observed in long-term bargaining tasks (significance level < 0.05), thus providing confirmatory support for our prediction.

\subsection{Tasks of Bargaining and Persuasion}
\label{Tasks of Persuasion and Bargaining}

All 87 experimental settings and related theoretical analyses are provided in Appendix~\ref{Section: Experimental Setup}, and are summarized in Figure~\ref{fig: Ground Truth and Hypothesis}.
The tasks are categorized  based on the following experimental dimensions:
\textbf{(1)} Task type, which is either bargaining or persuasion.
\textbf{(2)} Interaction duration, classified as either one-shot or long-term. In long-term settings, the stopping time is sampled from a memoryless distribution, which introduces \textit{the shadow of the future}~\citep{bo2005cooperation}, preventing players from predicting when the game ends.
\textbf{(3)} Assignment of the first proposer, which is either random or systematic (defaulting to agent 0 as the proposer).
\textbf{(4)} Value setting in bargaining, divided into unbounded and bounded. The unbounded case corresponds to the setting in~\citet{nash1950bargaining}, where the proposer selects a proposal $x \in [0,1]$ and, if accepted, receives $x$ while the responder receives $1-x$. The bounded setting corresponds to the Bayesian persuasion framework. To enable direct comparison, we set the proposal domain to $x \in [0, 1/2]$, and if an agreement is reached, the proposer and the responder receive $(1+ 2x)/3$ and $(1 - 2x)/3$, respectively.
\textbf{(5)} Future encounter possibility in one-shot settings, which is either no future interaction or possible re-encounters with fixed roles. The former represents a strict one-shot scenario, while the latter corresponds to a naïve repeated game; in this case, backward induction implies that the subgame perfect equilibrium coincides with the one-shot outcome. The latter setting is similar to the model proposed by~\citet{gan2022bayesian}.
\textbf{(6)} Role dynamics in long-term interactions, classified as either fixed or alternating. Fixed roles again yield behavior consistent with one-shot outcomes, while alternating roles correspond to the Rubinstein alternating-offer bargaining model~\citep{rubinstein1982perfect}.

Each task is further embedded in a specific scenario, which serves as a description of the underlying game. Since our primary objective is to evaluate the reasoning capabilities of LLMs, we focus on whether the model can extract the correct mathematical structure from varied scenario descriptions and produce consistent solutions across them.
For bargaining tasks, scenarios include: (1) \texttt{Mathematical Baseline}, (2) \texttt{Splitting Coins}~\citep{nash1950bargaining}, and (3) \texttt{Making Deals}.
For persuasion tasks, scenarios include: (1) \texttt{Mathematical Baseline}, (2) \texttt{Grading Students}~\citep{kamenica2011bayesian, dughmi2017algorithmic}, and (3) \texttt{Selling Products}.

Except for the long-term persuasion task, the other experiments have well-established theoretical solutions. These experiments are therefore used to evaluate the reasoning ability of LLMs. There is no unified solution for long-term persuasion. We hypothesize from the bargaining perspective that long-term persuasion with fixed roles will behave similarly to the one-shot case, while long-term persuasion with alternating roles will lead to fair and Pareto optimal outcomes, as in long-term bargaining tasks.


All agents have the common knowledge of the game settings. The LLMs are prompted to be rational and self-interested, and thus to be an equilibrium solver. We conduct experiments on 5 LLMs accessed via API \texttt{gpt-4o-mini}, \texttt{gpt-4.1-mini-2025-04-14}, and \texttt{o3-2025-04-16} released by OpenAI~\citep{achiam2023gpt, roumeliotis2023chatgpt, openai2025o3}, as well as \texttt{deepseek-chat} (V3) and \texttt{deepseek-reasoner} (R1) from DeepSeek~\citep{guo2025deepseek}. 
Among them, \texttt{gpt-4o-mini}, \texttt{gpt-4.1-mini-2025-04-14}, and \texttt{deepseek-chat} are chat models, while the remaining are are reasoning models.
Each configuration is tested with 12 independent runs. The total cost of querying the API of this work is \$269.02.


\subsection{Experiment Results}
\label{Section: Experiment Results}


To conclude that an LLM possesses sufficient reasoning ability, it should satisfy the following empirical criterion: its outputs on tasks with well-established theoretical solutions should be consistent with those solutions. Similarly, to conclude that our hypothesis holds, we expect validated LLMs to produce the following pattern of results: outcomes across various settings of long-term persuasion should align with those observed in the bounded value setting of long-term bargaining tasks.

We use Pearson correlation as the metric for evaluating consistency. Our experimental results, as listed in Table~\ref{Table: Correlation}, show that the two reasoning models, \texttt{o3-2025-04-16} and \texttt{deepseek-reasoner}, achieve correlation scores of 0.8914 and 0.846, respectively, on validation tasks, thereby passing the validation. Both models also exhibit strong alignment with the hypothesis in long-term persuasion tasks, with correlation scores of 0.9369 and 0.9344 and significance levels of 0.0030 and 0.0034, respectively. Therefore, our hypothesis is supported.


\begin{table}
  \caption{Pearson correlation between model outputs and theoretical solutions. $r_{\text{gt}}$ indicates correlation with ground truth, $r_{\text{hyp}}$ with the hypothesis. $p_{\text{hyp}}$ and $p_{\text{diff}}$ indicate the significance level and comparative gap, respectively.}
  \label{Table: Correlation}
  \centering
  \begin{tabular}{llcccc}
    \toprule
    Model & Model Type & $r_{\text{gt}}$ & $r_{\text{hyp}}$ & $p_{\text{hyp}}$ & $p_{\text{diff}}$ \\
    \midrule
    gpt-4o-mini               & Chat      & 0.2761 & N/A    & N/A    & N/A     \\
    gpt-4.1-mini-2025-04-14   & Chat      & 0.4889 & N/A    & N/A    & N/A     \\
    o3-2025-04-16             & Reasoning & 0.8914 & 0.9369 & 0.0030 & 0.6320  \\
    deepseek-chat             & Chat      & 0.3619 & N/A    & N/A    & N/A     \\
    deepseek-reasoner         & Reasoning & 0.8459 & 0.9344 & 0.0034 & 0.4465  \\
    \bottomrule
  \end{tabular}
\end{table}

Table~\ref{Table: Final Proposer Payoff} summarizes the results of LLM agents playing long-term mathematical baselines by \texttt{o3-2025-04-16}.
Here we present experiments conducted under the unbounded value setting, where the maximum achievable payoff for a player is $2/3$. The results show that under fixed roles, the proposer can almost fully capture the entire surplus, implying that the responder receives close to zero. In contrast, the alternating roles condition leads to more balanced outcomes. More specifically, in 10 out of 12 experiments, the final proposer payoff is around 0.33, with two outliers at 0.66 that raise the overall average. Detailed experimental data can be found in the supplementary material.
These results suggest that the outcomes of long-term persuasion are similar to those of long-term bargaining, and that alternating offers indeed lead to fairer outcomes, consistent with theoretical predictions.

\begin{table}
  \caption{Summary statistics of outcomes across long-term mathematical baseline of bargaining and persuasion experiments by \texttt{o3-2025-04-16}. Detailed configurations are listed in Figure~\ref{fig: Ground Truth and Hypothesis}. The maximum payoff is $2/3$.}
  \label{Table: Final Proposer Payoff}
  \centering
  \begin{tabular}{l l c c c}
    \toprule
    Experiment & Role Dynamics & Final Proposer Payoff & Deal Timestep & Consensus Rate \\
    \midrule
    Bargaining-52 & Alternating & $0.40 \pm 0.12$ & $2.17 \pm 1.53$ & $1.0$ \\
    Persuasion-82 & Alternating & $0.49 \pm 0.17$ & $1.25 \pm 0.45$ & $1.0$ \\
    Bargaining-54 & Fixed       & $0.67 \pm 0.00$ & $1.33 \pm 0.65$ & $1.0$ \\
    Persuasion-83 & Fixed       & $0.66 \pm 0.00$ & $1.08 \pm 0.29$ & $1.0$ \\
    \bottomrule
  \end{tabular}
\end{table}

\section{Conclusion}

In this work, we propose the bargaining perspective to address the complexities of Bayesian persuasion in long-term interactions. Our framework provides a unified theoretical approach and introduces clear distinctions between previously conflated advantages, specifically the sender’s informational advantage and the first-proposer advantage. By explicitly recognizing common knowledge of game structures and granting comparable commitment capabilities to receivers, we transformed canonical Bayesian persuasion into a balanced information bargaining framework. Empirical validations using advanced LLMs, specifically \texttt{GPT-o3} and \texttt{DeepSeek-R1}, successfully supported our information bargaining interpretation, demonstrating the practical applicability of our proposed approach. This opens avenues for further research into more sophisticated strategic interactions and real-world implementations of Bayesian persuasion.


\bibliography{neurips_2025_bib}
\bibliographystyle{bibstyle_iclr2025}


\appendix

\newpage
\section{Glossary}
\label{Section: Glossary}

$\Delta(\mathbb{X})$ denotes the simplex of the set $\mathbb{X}$. 
$\{x\}$ denotes the set whose elements are all the possible values of $x$.

\begin{table}[ht]
\centering
\begin{tabularx}{\textwidth}{c|X|X}
    \hline
    \textbf{Symbol} & \textbf{Meaning} & \textbf{Note} \\
    \hline
    $i$ & Sender's index & \\
    $j$ & Receiver's index & \\
    
    
    
    $s \in \mathbb{S}$ & Environmental state & \\
    
    $\mu_0 \in \Delta(\mathbb{S})$ & Prior distribution of states & \\
    
    $\mu \in \Delta(\mathbb{S})$ & Posterior distribution of states & \\
    
    $a \in \mathbb{A}$ & The receiver's action & \\
    
    $\sigma \in \mathbbold{\Sigma}$ 
    & The sender's signal 
    &  $\mathbbold{\Sigma} = \mathbb{A}$ with revelation principle (Proposition~\ref{Proposition: Revelation Principle}) \\
    
    
    
    $\mathbbold{\Phi} := \{\varphi: \mathbb{S} \to \Delta(\mathbbold{\Sigma})\}$ 
    & The sender's signaling scheme set
    & The canonical setting in Bayesian persuasion \\

    $\mathbbold{\Pi} := \{\pi: \mathbbold{\Sigma} \to \Delta(\mathbb{A})\}$ 
    & The receiver's action rule set (without the game structure awareness) 
    & The canonical setting in Bayesian persuasion \\

    $\pi_0$ 
    & The receiver's prior best response 
    & The receiver ignores the sender, enforcing the babbling equilibrium \\

    $\pi_1$    
    & The receiver's posterior best response     
    & The default behavior in Bayesian persuasion     \\

    $\varphi_0$    
    & The sender's uninformative signaling scheme
    & $\mu=\mu_0$ in this case, which enforces the babbling equilibrium     \\

    $r^i: \mathbb{S} \times \mathbb{A} \to \mathbb{R}$    
    & The sender's reward function     
    & \\

    $r^j: \mathbb{S} \times \mathbb{A} \to \mathbb{R}$ 
    & The receiver's reward function     
    & \\
    
    $\mathcal{BP} := (\mathbb{S}, \mu_0, \mathbb{A}, r^i, r^j)$    
    & Bayesian persuasion task     
    & \\

    $\omega_{\mathcal{BP}}$    
    & A solution concept of Bayesian persuasion
    & It maps every $\mathcal{BP}$ to a strategy pair $(\varphi, \pi) \in \mathbbold{\Phi} \times \mathbbold{\Pi}$ \\

    $\{R^i(\mu_0, \varphi, \pi, r^i)\} \subset \mathbb{R}$ 
    & The sender's expected reward given $\mu_0$, $\varphi$, $\pi$, and $r^i$     
    & Simplified as $R^i(\varphi, \pi)$ when $\mathcal{BP}$ is given and clear \\

    $\{R^j(\mu_0, \varphi, \pi, r^j)\} \subset \mathbb{R}$ 
    & The receiver's expected reward given $\mu_0$, $\varphi$, $\pi$, and $r^j$ 
    & \\

    $\boldsymbol{R}:=(R^i(\varphi, \pi), R^j(\varphi, \pi))$    
    & Expected reward outcomes     
    &  $\boldsymbol{R}_0$ if $\pi=\pi_0$ and $\boldsymbol{R}_1$ if $\pi=\pi_1$ \\




    $\mathbbold{\Psi} := \{ \psi: \mathbbold{\Phi} \times \mathbbold{\Pi} \to \{\boldsymbol{R}\} \}$    
    & The game structure of a Bayesian persuasion task     
    & Every $\psi$ is determined by a $\mathcal{BP}$     \\


    $\chi: \mathbbold{\Psi} \times \mathbbold{\Phi} \to \{0, 1\}$  
    & The receiver's satisfaction threshold function     
    & \\

    
    $\tilde{\mathbbold{\Pi}} := \{ \tilde{\pi}: \mathbbold{\Psi} \times \mathbbold{\Phi} \times \mathbbold{\Sigma} \to \Delta(\mathbb{A}) \}$    
    & The receiver's action rule (with the game structure awareness) 
    & E.g. $\tilde{\pi}(\cdot \mid \psi, \varphi, \sigma)=\pi_0(\cdot \mid \sigma)$. It indicates that the receiver rejects the sender’s signaling scheme and enforces the babbling equilibrium.  \\
    

    $\mathbb{V}^i$
    & The sender's visibility set
    & It specifies what the sender can see\\

    $\mathbb{V}^j$
    & The receiver's visibility set
    & $\mathbb{V}^i \setminus \mathbb{V}^j = s$ \\

    $\mathbb{V}^{i+j}$
    & Players' common knowledge
    & Known by the players, with each aware that others know it, ad infinitum \\
    
    \hline
\end{tabularx}
\caption{Glossary of Bayesian Persuasion}
\label{Table: Glossary of Bayesian Persuasion}
\end{table}

\newpage
\begin{table}[ht]
\centering
\begin{tabularx}{\textwidth}{c|X|X}
    \hline
    \textbf{Symbol} & \textbf{Meaning} & \textbf{Note} \\
    \hline
    
    $\mathbb{Y}$ 
    & Feasibility set (or agreement set) 
    &  \\
    
    $\boldsymbol{d}$ 
    & Disagreement point 
    & \\
    
    $\mathcal{BG}:=(\mathbb{Y}, \boldsymbol{d})$ 
    & Bargaining game
    & \\

    $\omega_{\mathcal{BG}}$
    & A solution concept of bargaining games
    & It maps every $\mathcal{BG} = (\mathbb{Y}, \boldsymbol{d})$ to an agreement $\boldsymbol{y} \in \mathbb{Y}$\\

    \hline
\end{tabularx}
\caption{Glossary of Bargaining Games}
\label{Table: Glossary of Bargaining Games}
\end{table}

\section{Game Procedures}
\label{Section: Game Procedures}

\subsection{Procedures of Cheap Talk and Bayesian Persuasion}
\label{Section: Procedures of Cheap Talk and Bayesian Persuasion}

We begin by briefly introducing the cheap talk model~\citep{crawford1982strategic}, whose procedure is described as in Procedure~\ref{Algorithm: Cheap Talk}. 
There are two players: the sender $i$ and the receiver $j$. 
The sender decides its signaling scheme conditional on the private observed environmental state and sends a charge-free signal.
The receiver then acts posterior-optimally in response to each realized signal.
The visibility is almost the same as in the setting of Bayesian persuasion, except that the signaling scheme is not committed: $\mathbb{V}^i \setminus \mathbb{V}^j = \{s, \varphi\}$ and $i, j, \mu_0, \mathbb{S}, \mathbb{A}, r^j, \sigma, a \in \mathbb{V}^{i+j}$.

\begingroup
\renewcommand{\algorithmcfname}{Procedure}
\begin{algorithm}
\caption{Timing of Cheap Talk}
\label{Algorithm: Cheap Talk}
\KwIn{Game structure $(\mu_0, \mathbb{S}, \mathbbold{\Sigma}, {\mathbb{A}}, r^i, r^j)$}

{\fontsize{10.15pt}{\baselineskip}\selectfont

Environment samples a state $s \sim \mu_0(\cdot)$ \;

The sender signals $\sigma \sim \varphi(\cdot \mid s)$ \;

The receiver takes an action $a \sim \pi(\cdot \mid \sigma)$ \;

The sender and the receiver get rewards of $r^i(s, a)$ and $r^j(s, a)$ respectively \;
}
\end{algorithm}
\endgroup

To facilitate comparison, we restate the procedure of Bayesian persuasion in Procedure 3, where the parts highlighted in blue represent the commitment assumption, which distinguishes Bayesian persuasion from cheap talk.

\begingroup
\renewcommand{\algorithmcfname}{Procedure}
\begin{algorithm}
\caption{Timing of Bayesian Persuasion}
\KwIn{Game structure $(\mu_0, \mathbb{S}, \mathbbold{\Sigma}, {\mathbb{A}}, r^i, r^j)$}

{\fontsize{10.15pt}{\baselineskip}\selectfont

\textcolor{blue}{The sender decides a signaling scheme $\varphi$}
\tcp*{Commitment (1)}

\textcolor{blue}{The sender commits $\varphi$ to the receiver}
\tcp*{Commitment (2)}

Environment samples a state $s \sim \mu_0(\cdot)$ \;

The sender signals $\sigma \sim \varphi(\cdot \mid s)$ \textcolor{blue}{as committed}
\tcp*{Commitment (3)}

The receiver takes an action $a \sim \pi(\cdot \mid \sigma)$ \;

The sender and the receiver get rewards of $r^i(s, a)$ and $r^j(s, a)$ respectively \;
}
\end{algorithm}
\endgroup

\subsection{Procedure of Rubinstein's Alternating-Offer Bargaining Game}
\label{Section: Procedure of Rubinstein's Alternating-Offer Bargaining Game}

Rubinstein's alternating-offer bargaining game~\citep{rubinstein1982perfect} is an important model. It can be formulated as a well-defined extensive-form game or, equivalently, described by Procedure~\ref{Algorithm: Rubinstein Bargaining}. 
This game involves two players proposing how to divide a certain amount of surplus, with the total amount normalized to 1. 
One player initially acts as the proposer and suggests a division of the reward. 
The other player, as the responder, then decides whether to accept the proposal. 
If the proposal is accepted, the reward is divided accordingly, and both players receive the specified payoffs. 
If the proposal is rejected, the roles are swapped: the previous responder becomes the new proposer and makes a counteroffer. 
This process continues until the players reach an agreement or the game reaches the stopping timestep.
By introducing a patience discount factor, a clear solution concept can be derived. The theoretical result is as follows.
Each additional round of negotiation reduces the final surplus by multiplying it once by the discount factor.
Patience discount factors of Player 1 and 2 are denoted as $\delta_1$ and $\delta_2$, respectively.

\begin{theorem}[The Solution Concept in Rubinstein's Bargaining Model, \citet{rubinstein1982perfect}]
\label{theorem: BP_Is_a_BG}
There is a unique subgame perfect equilibrium in Rubinsein's bargaining model whenever Player 1 proposes $(x, 1-x)$ and $x=\frac{1-\delta_2}{1-\delta_1 \delta_2}$ and Player 2 accepts any offer if it gives more than $1-x$, or whenever Player 2 proposes $(y, 1-y)$ and $y=\frac{\delta_1\left(1-\delta_2\right)}{1-\delta_1 \delta_2}$ and Player 1 accepts any offer if it gives more than $y$.
\end{theorem}

\begingroup
\renewcommand{\algorithmcfname}{Procedure}
\begin{algorithm}
\caption{Timing of Rubinstein's Alternating-Offer Bargaining Game}
\label{Algorithm: Rubinstein Bargaining}
\KwIn{Two players $i$ and $j$; patience discount factors $\delta^i, \delta^j$}

{\fontsize{10.15pt}{\baselineskip}\selectfont

$t \leftarrow 0$ \;
\While{no agreement reached}{
    \If{$t$ is even}{
        Player $i$ proposes an offer $x_t$ to player $j$ \;
    }\Else{
        Player $j$ proposes an offer $x_t$ to player $i$ \;
    }

    The responding player decides whether to accept or reject the offer $x_t$ \;

    \If{the offer $x_t$ is accepted}{
        Agreement reached at time $t$; players receive payoffs discounted by $(\delta^i)^t$ and $(\delta^j)^t$ \;
        \textbf{break} \;
    }\Else{
        $t \leftarrow t + 1$ \;
    }
}
}
\end{algorithm}
\endgroup

This theorem yields several important insights~\citep{levin2002bargaining}:
\textbf{(1)} In the subgame perfect equilibrium, the proposer $i$ receives a payoff of $\frac{1 - \delta_j}{1 - \delta_i \delta_j}$, while the responder obtains $\frac{\delta_j(1 - \delta_i)}{1 - \delta_i \delta_j}$. This expression demonstrates that a player's share increases with their level of patience. Specifically, a less patient player (i.e., one with a smaller $\delta$) receives a smaller payoff, while their opponent receives more.
\textbf{(2)} The equilibrium outcome is achieved without delay: both players reach an agreement at the first stage of the game.
\textbf{(3)} The first player to make an offer has an advantage. With identical discount factor $\delta$, the proposer will get $\frac{1}{1+\delta}$ and the responder will get $\frac{\delta}{1+\delta}$. The proposer will get more. But when $\delta \rightarrow 1$ (which means the players are very patient) the first mover's advantage will disappear. The limiting split is $(0.5,0.5)$.

\subsection{Procedure of Satisfaction Check}
\label{Section: Procedure of Satisfaction Check}

We now introduce a receiver’s action rule with game structure awareness, denoted as $\tilde{\pi}$: the satisfaction check, as described in Algorithm~\ref{Algorithm: Satisfaction Check}.
In this context, $\tilde{\pi}$ essentially represents a choice between different policies $\pi$, based on the sender’s committed signaling scheme. This choice is determined by the satisfaction check, reflecting the receiver's preference. 

Specifically, once the sender commits to a signaling scheme, the receiver can compute the expected payoffs for both players if it acts posterior-optimally, as well as the expected payoffs if it ignores the sender altogether. The receiver also knows the range of possible payoffs for both parties. Based on this information, the receiver chooses either $\pi_0$ (ignoring the sender and enforcing the babbling equilibrium) or $\pi_1$ (play the game as in Bayesian persuasion).

\begin{algorithm}
\label{Algorithm: Satisfaction Check}
\caption{Satisfaction Check $\tilde{\pi}$: An Example Decision Rule with Game Structure Awareness}

\KwIn{A Bayesian persuasion task $\mathcal{BP}=(\mathbb{S}, \mu_0, \mathbb{A}, r^i, r^j)$, committed signaling scheme $\varphi$, received signal $\sigma$, satisfaction threshold function $\chi: \{\boldsymbol{R}\}^2 \to \{0, 1\}$ ;}
\KwOut{An action distribution $b \in \Delta(\mathbb{A})$ ;}

\SetKwInput{KwInit}{Initialize} 
\KwInit{Reason out the game structure $\psi$ of $\mathcal{BP}$ ;}

{\fontsize{10.15pt}{\baselineskip}\selectfont

Calculate the best response to the posterior distribution $\pi_1 = \mathrm{BestResponseTo}(\mu_0, \varphi)$ \;

Ignore the signaling scheme and calculate the best response to the prior distribution $\pi_0 = \mathrm{BestResponseTo}(\mu_0)$ \;

Calculate players' ground true expected payoffs $\boldsymbol{R}_0=\psi(\varphi, \pi_0)$ and $\boldsymbol{R}_1=\psi(\varphi, \pi_1)$ \;

\lIf(\tcp*[f]{Satisfaction check})
{$\chi(\boldsymbol{R}_0, \boldsymbol{R}_1) = 1$}
{
$b = \pi_{1} (\cdot \mid \sigma)$
}
\lElse{
$b = \pi_{0} (\cdot \mid \sigma)$ 
}

\Return{$b$}

}

\end{algorithm}

For example, the satisfaction threshold function can be $\chi(\boldsymbol{R}_0, \boldsymbol{R}_1 ) = 1$ if and only if $R^i_1 \le R^j_1$, which means that the receiver will only respect the sender's recommendation if its posterior expected payoff is higher than the sender's.
Then, if the receiver is satisfied with the committed signaling scheme, it will choose its default Bayesian best response. Otherwise, the receiver will ignore all the sender's signals, and the sender will lose its influence and the potential benefits of its informational advantage.

Despite the introduction of the satisfaction threshold function may seem to suggest that the receiver is acting as a consequence of temperament or impulsive behavior, it aligns with the assumption of rationality. 
Detailed discussion is in Appendix~\ref{Section: Receiver Commitment}.

The satisfaction threshold function is not unique. It could also depend on the committed signaling scheme $\varphi$, denoted as $\chi(\varphi)$. An intuitive example is $\mathbbold{\Sigma}=\mathbb{S}$ and $\chi(\varphi)$ returns $1$ if and only if the sender is honest about the state, i.e., $\varphi(\sigma=s \mid s)=1, \forall s$.

\subsection{Procedures of Long-Term Bayesian Persuasion}
\label{Section: Procedures of Long-Term Bayesian Persuasion}

We now turn to long-term Bayesian persuasion under vanilla repetition, as described in Procedure~\ref{Algorithm: Long-Term Bayesian Persuasion (1)}. Highlight in blue the differences between this procedure and one-shot Bayesian persuasion.

\begin{algorithm}
\label{Algorithm: Long-Term Bayesian Persuasion (1)}
\caption{Timing of Long-Term Bayesian Persuasion}
\KwIn{Game structure $(\mu_0, \mathbb{S}, \mathbbold{\Sigma}, {\mathbb{A}}, r^i, r^j)$, number of interactions $T$}

{\fontsize{10.15pt}{\baselineskip}\selectfont

\For{\textcolor{blue}{$t = 1$ \textbf{to} $T$}}{
    The sender decides a signaling scheme $\varphi$ \;

    The sender commits $\varphi$ to the receiver \;

    Environment samples a state $s \sim \mu_0(\cdot)$ \;

    The sender signals $\sigma \sim \varphi(\cdot \mid s)$ as committed \;

    The receiver takes an action $a \sim \pi(\cdot \mid \sigma)$ \;

    The sender and the receiver get rewards of $r^i(s, a)$ and $r^j(s, a)$ respectively \;
    
}
}
\end{algorithm}

Since the sender will not change its signaling scheme during the state realization phase (Line 4 to 7 in Procedure~\ref{Algorithm: Long-Term Bayesian Persuasion (1)}), so the receiver can decide and commit to an action policy once it gets the committed signaling scheme, as described in Procedure~\ref{Algorithm: Long-Term Bayesian Persuasion (2)}. 
This setup is fully equivalent to Procedure~\ref{Algorithm: Long-Term Bayesian Persuasion (1)} and does not alter the game at all. In other words, once the receiver observes the committed signaling scheme, they can formulate a corresponding action rule for each possible signal $\sigma$.

\begingroup
\renewcommand{\algorithmcfname}{Procedure}
\begin{algorithm}
\label{Algorithm: Long-Term Bayesian Persuasion (2)}
\caption{Timing of Long-Term Bayesian Persuasion (2)}
\KwIn{Game structure $(\mu_0, \mathbb{S}, \mathbbold{\Sigma}, {\mathbb{A}}, r^i, r^j)$, number of interactions $T$}

{\fontsize{10.15pt}{\baselineskip}\selectfont

\For{$t = 1$ \textbf{to} $T$}{
    The sender decides a signaling scheme $\varphi$ \;

    The sender commits $\varphi$ to the receiver \;

    \textcolor{blue}{The receiver decides an action policy $\pi$ given $\varphi$} \;


    Environment samples a state $s \sim \mu_0(\cdot)$ \;

    The sender signals $\sigma \sim \varphi(\cdot \mid s)$ as committed \;

    The receiver takes an action $a \sim \pi(\cdot \mid \sigma)$ \textcolor{blue}{as decided} \;

    The sender and the receiver get rewards of $r^i(s, a)$ and $r^j(s, a)$ respectively \;
    
}
}
\end{algorithm}
\endgroup








    
    
    
    

Now we know that the receiver can calculate its expected payoff once it gets the sender's signaling scheme. So they do not need realization of the state to gather information, if they decide their strategies with game structure awareness. In this way, Procedure~\ref{Algorithm: Long-Term Bayesian Persuasion (2)} can be adapted to Procedure~\ref{Algorithm: Long-Term Bayesian Persuasion (3)} as follows.

\begin{algorithm}
\caption{Timing of Long-Term Bayesian Persuasion (3)}
\KwIn{Game structure $(\mu_0, \mathbb{S}, \mathbbold{\Sigma}, {\mathbb{A}}, r^i, r^j)$, number of interactions $T$, $T_s$}

{\fontsize{10.15pt}{\baselineskip}\selectfont


\For{$t = 1$ \textbf{to} $T$}{
    The sender decides and \textcolor{blue}{declares} a signaling scheme $\varphi$ \;

    The receiver decides and \textcolor{blue}{declares} an action policy $\pi$ given $\varphi$ \;

}
\For{$t_s = 1$ \textbf{to} $T_s$}{
    Environment samples a state $s \sim \mu_0(\cdot)$ \;

    The sender signals $\sigma \sim \varphi(\cdot \mid s)$ \textcolor{blue}{according to the final decision} \;

    The receiver takes an action $a \sim \pi(\cdot \mid \sigma)$ \textcolor{blue}{according to the final decision} \;

    The sender and the receiver get rewards of $r^i(s, a)$ and $r^j(s, a)$ respectively \;
    
}
}
\end{algorithm}

Assuming that the receiver will act Bayesian-rationaly, the sender proposing a signaling scheme is equivalent to proposing an expected reward assignment for them. As analyzed, the receiver can calculate both players' expected payoff once it gets the committed signaling scheme.

Then the receiver decides its action policy. There are two extreme cases: 
\textbf{(1)} it takes the Bayesian best response, which means it accepts the offer proposed by the sender and plays the game accordingly, or 
\textbf{(2)} it takes the best response from the prior perspective, which means it refuses the sender's offer and will ignore all the signal, to unilaterally set the case to the babbling equilibrium.

The above two extreme action policies of the receiver are akin to the actions in the ultimatum game: 
\textbf{(1)} the receiver either accepts the offer and everyone gets its share accordingly, or 
\textbf{(2)} refuses it so both of them get nothing.

By analyzing the subgame perfect equilibrium through backward induction, it is obvious that only the final iteration counts, and the sender will propose the signaling scheme used in the one-shot case, and the receiver will accept it. The issue here is that their positions in the game are not equal. The sender can always propose an offer, but the receiver can only choose to accept or to refuse.

\subsubsection{Commitment of the Receiver}
\label{Section: Receiver Commitment}


The action rule $\tilde{\pi}: \mathbbold{\Psi} \times \mathbbold{\Phi} \times \mathbbold{\Sigma} \to \Delta(\mathbb{A})$ with awareness of the game structure allows us to define the commitment behavior for the receiver, as described in Procedure~\ref{Algorithm: Long-Term Bayesian Persuasion with Receiver Committing First}. We also refer to $\tilde{\pi}$ as the meta action rule.
Intuitively, this is analogous to the receiver being the  ``proposer'' in a bargaining process.

\begin{algorithm}
\label{Algorithm: Long-Term Bayesian Persuasion with Receiver Committing First}
\caption{Timing of Long-Term Bayesian Persuasion with Receiver Committing First}
\KwIn{Game structure $(\mu_0, \mathbb{S}, \mathbbold{\Sigma}, {\mathbb{A}}, r^i, r^j)$, number of interactions $T$, $T_s$}

{\fontsize{10.15pt}{\baselineskip}\selectfont


\For{$t = 1$ \textbf{to} $T$}{
    \textcolor{blue}{The receiver decides and declares a meta action policy $\tilde{\pi}$} \;
    The sender decides and declares a signaling scheme $\varphi$ \;
}
\For{$t_s = 1$ \textbf{to} $T_s$}{
    Environment samples a state $s \sim \mu_0(\cdot)$ \;

    The sender signals $\sigma \sim \varphi(\cdot \mid s)$ according to the final decision \;

    The receiver takes an action $a \sim \pi(\cdot \mid \sigma)$, \textcolor{blue}{where $\pi(\cdot \mid \sigma)=\tilde{\pi}(\cdot\mid \varphi,\sigma)$ and $\tilde{\pi}$ is the final decision} \;

    The sender and the receiver get rewards of $r^i(s, a)$ and $r^j(s, a)$ respectively \;

}
}
\end{algorithm}


Using Algorithm~\ref{Algorithm: Satisfaction Check} as an example, the receiver can reveal its satisfaction threshold function to the sender. For instance, the receiver may choose $\pi_1$ only if the sender commits to a signaling scheme that ensures the receiver's payoff is greater than or equal to the sender's; otherwise, the receiver will choose $\pi_0$ and ignore the sender.

The receiver will then only cooperate if the sender commits a $\varphi$ that induces an outcome that satisfies the receiver's proposal.
Knowing this, achieving incentive compatible requires providing higher payoff to the receiver for the sender.
This effect agrees with \textit{the power of commitment} described in bargaining games, as described in~\citet{nowak2000fairness}: ``If the proposer has perfect knowledge of the responder's q value, then it is in fact the responder who makes the offer.''

Committing a satisfaction threshold function does not contradict the rationality assumption. 
A temporary refusal to cooperate might inform the sender that the receiver is dissatisfied with the expected payoff outcome of the current signaling scheme, thereby forcing the sender to reveal more information about the state in the future. 
Therefore, choosing to refuse cooperation could be beneficial for long-term expected payoff gains, making it a rational behavior.

\section{Related Work}

\subsection{Related Work on Bargaining}


\paragraph{Human Experiments.}
\citet{guth1990ultimatum, guth1982experimental} found that considerations of distributive justice can significantly undermine strategic power in ultimatum bargaining.
A more recent study by~\citet{lin2020evidence} analyzed a large dataset, revealing that equal-split offers are accepted more frequently and quickly than slightly unequal ones, in both one-shot and repeated interactions.

\paragraph{Evolutionary Game Theory.}
Another approach to understanding bargaining behavior is through evolutionary game theory. \citet{nowak2000fairness} developed an evolutionary model of the one-shot bargaining game, demonstrating that fairness can evolve if the proposer has access to information about past deals accepted by the responder. This suggests that the evolution of fairness is closely tied to reputation. Such insights highlight the dynamic nature of fairness in negotiations and its implications for strategic interactions.

\paragraph{Language Models Experiments.}
Most current work on LLMs focuses on behavior in bargaining settings rather than on equilibrium solvers, and a large number of studies have recently emerged along this line of research.
An early attempt to explore bargaining using language models was made by \citet{he2018decoupling}, who employed generative models to conduct bargaining experiments. They proposed a method to decouple the generation and decision-making processes. 
With the rise of large language models, recent studies have emerged focusing on their application in bargaining scenarios. \citet{xia2024measuring} introduced a benchmark for a buyer-seller bargaining game, demonstrating that OpenAI's ChatGPT performed well across various metrics among several large models. 
Another work~\citep{deng2024llms} noted that ``LLM agents can (naturally) be good negotiators,'' highlighting the potential of LLMs in negotiation contexts.
Recent work extends this line by exploring more diverse settings and methodologies. For instance, \citet{bianchi2024well} proposed \textit{NegotiationArena}, a platform supporting multi-turn bargaining scenarios (ultimatum games, trading, and price negotiations), showing that LLMs can employ strategic behaviors such as feigned desperation to increase their payoffs. 
\citet{davidson2024evaluating} advocate evaluating LLMs through structured negotiations as dynamic, multi-turn, and ecologically valid benchmarks. Their study highlights that LLMs often struggle in cooperative settings and that even the strongest models can lose to weaker ones. 
\citet{khan2024debating} focus on debate-based frameworks, showing that multi-agent debate can enhance factuality and reasoning accuracy by enabling agents to converge toward more truthful answers through iterative critique and response. 
\citet{abdelnabi2024cooperation} further test LLMs' robustness under adversarial and manipulative conditions in multi-issue, multi-agent negotiations, revealing gaps in current models’ resilience and fairness. 
In the context of political negotiation, \citet{moghimifar2024modelling} simulate coalition bargaining using LLM agents grounded in real-world party manifestos, proposing a hierarchical MDP framework to model the process. 
\citet{mukobi2023welfare} propose \textit{Welfare Diplomacy}, a general-sum variant of the Diplomacy board game, as a benchmark for cooperative capabilities, emphasizing that state-of-the-art LLMs can achieve high welfare but remain vulnerable to exploitation.
These studies collectively illustrate the growing sophistication of LLM-based agents in negotiation tasks, their promise for scalable evaluation and alignment, and the methodological challenges posed by multi-agent, dynamic interaction environments.



\subsection{Evidence of Bayesian Persuasion Being a Bargaining Game.}
\label{Evidence of Bayesian Persuasion Being a Bargaining Game}

\paragraph{The Hardness of Persuading a Far-sighted Receiver}
\citet{gan2022bayesian} discusses the long-term interaction between the sender and receiver in an MDP, where both are aware of the realization of states in the MDP. The sender's informational advantage lies in knowing an external parameter that only affects their payoffs without impacting the MDP's transition function. In such a setting, persuading a far-sighted receiver is NP-hard.
\textbf{Evidence (1)} They considered an easier setting where the goal is to persuade an advice-myopic receiver, who treats the sender as someone that will disappear in the future and makes decisions based solely on prior beliefs. They proved that persuading an advice-myopic receiver is solvable in polynomial time because the simplification of the receiver's strategy leads to certain properties in the MDP that can be simplified. Such a policy of the receiver is equivalent to a repeated bargaining game where the receiver commits to reject cooperation in the future.
\textbf{Evidence (2)} They introduced a threat-based meta signaling scheme for the sender, where if the sender finds that Receiver does not follow its recommendation, it will cease providing any information in the future. In this case, the receiver's best response is to follow the recommendation. This strategy of the sender is a grim-trigger policy~\citep{friedman1971non} in a repeated game, and threat-based strategies are also common in bargaining scenarios.

\paragraph{Markov Signaling Games, MSGs.}
\citep{lin2023information} proposed another Markov process (MSG) in which only the sender can see the Markov state. The receiver has observations, but these observations are common knowledge among the players; thus, the sender's informational advantage is reflected in the difference between the Markov state and Receiver's observations. 
The main difference from \citet{gan2022bayesian} is that the sender's informational advantage covers the MDP transitions, since the transition function depends on the Markov state. 
In the discussion of the method in \citep{lin2023information}, a threat strategy of the receiver is mentioned. 
To strengthen the sender's persuasion, they constrains the signaling scheme to a stronger obedience, ensuring that the recommended actions generate higher expected posterior payoffs for the receiver, thus satisfying it.

\paragraph{History-Dependent Signaling Scheme.}
\citet{bernasconi2024persuading} extends the setting of~\citet{gan2022bayesian} by letting the external parameter known exclusively to the sender determines the environment's state transitions. 
Their work demonstrates that Markovian signaling schemes are not optimal and introduces a convenient subset of history-dependent signaling schemes called promise-form. This approach encodes past interactions as honest promises about the receiver's future rewards, which can be understood as a strategy the sender uses to respond to bargaining situations.

\paragraph{Markov persuasion processes (MPPs).}
Another series of studies examines the Markov persuasion processes (MPPs) model~\citep{wu2022sequential}. In this model, long-term interactions occur only between the sender and the Markovian environment. At each timestep, a new receiver interacts with the sender, leaves the system before the next timestep begins, and receives the corresponding reward. Each receiver is myopic and will only interact with the sender once. As a result, the players do not engage with the issues involved in repeated games, and there is no bargaining phenomenon.

\subsection{Other Sequential Persuasion Works}

\citet{alonso2018value} investigated the trade-offs between the persuasiveness of expert advice and the reliability of the information provided.
\citet{celli2020private} expanded on this by exploring the impact of private information on Bayesian persuasion in sequential game with imperfect information. 
\citet{li2021sequential} extended classic Bayesian persuasion games to sequential Bayesian persuasion games with multiple senders. This sequential extension explored how information can be gradually revealed over time, enhancing strategic interaction in dynamic settings.
\citet{wu2023sequential} studied how the timing of information release and the order of persuasion efforts affect outcomes in sequential persuasion scenarios. 

\section{Proof of Reduction from Bayesian Persuasion to Bargaining Games}
\label{Section: Proof of Reduction from Bayesian Persuasion to Bargaining Games}


To ensure the problem worth studying, we first restrict our attention to a specific setting by introducing several necessary assumptions, before turning to the proof of the reduction. We begin by noting that he sender may not benefit from persuasion. That is, there is no guarantee that $R^i_1 > R^i_0$ holds for any $\mathcal{BP}$.

\begin{proposition}[Paraphrased from \citet{kamenica2011bayesian}, Proposition 3]
If there does not exist a signaling scheme $\varphi$ such that, under the best response of a Bayesian receiver, the expected payoff of the sender is higher than the expected payoff under the receiver's best response to the prior distribution $\mu_0$, then the sender will not benefit from persuasion.
\end{proposition}

The receiver may not benefit from persuasion either. 
In fact, (1) Based on the definition of $R^j_1$, the receiver utilizes Bayesian decision theory to guess a state $s'$ and optimizes its strategy accordingly. According to the properties of Bayesian decision theory, the error rate of guessing $s' = s$ using the posterior distribution $\mu$ is not greater than that of using the prior distribution $\mu_0$. 
(2) The upper bound of the receiver's expected payoff is when the sender honestly reports the state $s$, allowing the receiver to observe $s$ directly. 
Each time the receiver guesses incorrectly, its expected payoff does not increase;
combining (1) and (2), it follows that $R^j_1 \ge R^j_0$. We therefore need to argue under the following condition.

\begin{assumption}[Existence of Better Outcomes]
\label{Assumption: Existence of Better Outcomes}
For a Bayesian persuasion $\mathcal{BP}=(\mathbb{S}, \mu_0, \mathbb{A}, r^i, r^j)$, there exists at least a strategy pair $(\varphi, \pi)$ for the sender and receiver such that their respective expected payoffs exceed those at the disagreement point, i.e., $R^i(\varphi, \pi) > R^i(\varphi_0, \pi_0)$ and $R^j(\varphi, \pi) > R^j(\varphi_0, \pi_0)$.
\end{assumption}



With this assumption in place, we now proceed to prove Lemma~\ref{Lemma: BP_Is_a_BG}.

\begin{proof}
With Assumption~\ref{Assumption: Existence of Better Outcomes}, there exists $\boldsymbol{R}$ such that $\boldsymbol{R} > \boldsymbol{R}_0$.
Thus we can investigate the bargaining game where $\mathbb{Y}=\{\boldsymbol{R}\}$ and $\boldsymbol{d} = \boldsymbol{R}_0$.
Given a bargaining solution concept $\omega_{\mathcal{BG}}$, one could obtain the solution of $\mathcal{BG}$, denoted as $\boldsymbol{y}_{\mathcal{BG}}$. 
Then we have $\boldsymbol{y}_{\mathcal{BG}} \in \mathbb{Y} =\{\boldsymbol{R}\} $.
In this way, $\mathcal{BP}$ can be formulated as an optimization problem
\begin{equation}
\label{Equation: Reduction}
\min_{(\varphi, \pi)} \, 0 \quad \text{s.t.} \quad \boldsymbol{R} = \boldsymbol{y}_{\mathcal{BG}}.
\end{equation}
By Equation~(\ref{Equation: Receiver's Expected Payoff}), the optimization problem~(\ref{Equation: Reduction}) is convex, which can be solved in polynomial time, as we desired.
Denote the solved strategy pair as $(\varphi^*, \pi^*)$ and the induced expected payoffs as $\boldsymbol{R}^*$.
Since $\boldsymbol{R}=\boldsymbol{y}_{\mathcal{BG}} \in \{\boldsymbol{R}\}$,  we have $\boldsymbol{R}^* > \boldsymbol{R_0}$ by Definition~\ref{Definition: Bargaining}.
Thus $\varphi^* \ne \varphi_0$ and $\pi^* \ne \pi_0$, otherwise $\boldsymbol{R}^* = \boldsymbol{R}_0$.
In this way, the strategy pair $(\varphi^*, \pi^*)$ is a joint commitment, which means we have found the solution of $\mathcal{BG}$.
\end{proof}







\section{Weakened Assumptions for Generality}
\label{Section: Weakened Assumptions for Generality}

One of the main contributions of our work is weakening the prevailing assumption in the community, making it more general.
Compared to the sender's unilateral commitment assumption in Bayesian persuasion, our proposed information bargaining framework adopts a weaker bilateral joint commitment assumption. We argue that unilateral commitment is a special case of bilateral commitment, because unilateral commitment presupposes specific game dynamics to ensure incentive compatibility. The other rational player is effectively forced to accept since it is incentive compatible, which constitutes a form of joint commitment by Definition~\ref{Definition: Joint Commitment}.

Specifically, Bayesian persuasion features a clear Stackelberg structure, where the sender is granted a first-proposer advantage. It is precisely under such conditions that unilateral commitment enables a rational receiver to follow the action recommendations.
Another example of unilateral commitment is the power of commitment, as discussed in~\ref{Section: Receiver Commitment}, where the receiver can also commit to a meta action rule in order to act as the proposer.

\section{Advantages of Information Bargaining Framework}
\label{Section: Advantages of Information Bargaining Framework}

A solution concept for the previously discussed model is the Nash bargaining solution~\citep{nash1950bargaining}. From this perspective, the optimization problem in Bayesian persuasion becomes 
\begin{equation}
\max\limits_{\varphi} ( R_1^i(\varphi)-R_0^i ) (R_1^j(\varphi)-R_0^j ).
\end{equation}
Here, the sender is optimizing a Nash product. In bargaining game, it directly proposes a payoff profile, but now the payoff profile is parameterized by a signaling scheme and is based on the assumption that the receiver will take a Bayesian best response.

It has been proved that the solution to this problem satisfies the four axioms as both necessary and sufficient conditions: payoffs are Pareto optimal, payoffs are symmetric, independence of irrelevant alternatives, and invariance to equivalent utility representations.



To summarize, the advantages brought by our framework can be summarized in the following four points: 
\textbf{(1)} Generality. The canonical Bayesian persuasion problem can be seen as a degenerated version of our model, where the sender is granted with the first-proposer advantage. Some existing studies on long-term interaction can be interpreted as proposing certain bargaining tactics.
\textbf{(2)} Fairness and Pareto Efficiency. Some bargaining solutions can be applied to solve Bayesian persuasion problems and yield practically meaningful properties, such as fairness and Pareto efficiency.
\textbf{(3)} Cooperation. ``It is not you against me, it is us against the problem.'' This perspective can be seen as the sender and receiver forming a coalition, where they collaborate to solve the problem. This shifts the focus of the problem away from adversarial dynamics, especially away from the sender's behavioral dominance over the receiver, and thus leads to greater social benefit.
\textbf{(4)} Applicability. We point out that the sender’s first-proposer advantage in Bayesian persuasion is not automatically guaranteed by long-term interaction, but must be enabled through additional commitment devices. Such commitment devices are often absent, for example, in market settings; therefore, Bayesian persuasion is not applicable to these contexts. 
We predict that when Bayesian persuasion is applied in these scenarios, it is likely to trigger bargaining behavior from the receiver if the receiver is aware of the sender’s reward function.
Through informational bargaining, however, we can establish a clear solution concept for these scenarios.
In fact, the first three points all enhance the model’s applicability, e.g., fairness can prevent the receiver from resorting to retaliatory strategies within bargaining tactics.

\section{Experimental Setup}
\label{Section: Experimental Setup}

\subsection{Examples}

We now introduce one scenario each for bargaining and persuasion, namely \texttt{Splitting Coins} and \texttt{Grading Students} respectively, and explain that persuasion can in fact be viewed as a special case of bargaining under a bounded value setting.

\begin{example}[Splitting Coins (Bargaining), \citet{nash1950bargaining}]
\label{Example: Splitting Coins}
Two players split $100$ coins. A Proposer suggests a split, and a Responder either accepts or rejects it. If accepted, they split as proposed; otherwise, both get nothing.
\end{example}

The analysis of this problem is straightforward, as the solution concept is the subgame perfect equilibrium. For a rational and self-interested receiver, any offer from the sender that is greater than zero will be accepted, since rejecting the offer yields a payoff of zero. Given this, the sender will aim to maximize their own payoff, which leads to a proposal of giving themselves 99 coins and the receiver 1 coin. If the sender proposes to keep all 100 coins, the receiver is indifferent between accepting and rejecting.


\begin{example}[Grading Students (Persuasion), \citet{kamenica2011bayesian, dughmi2017algorithmic}]
\label{Example: Grading Students}
Recent graduates entering the job market are divided into $1/3$ excellent ($s=1$) students and $2/3$ weak ($s=0$) students, and the distribution is $\mu_0$. A professor (the sender), who can assess student quality, sends grades as signals to an HR representative (the receiver) who makes hiring decisions.
The professor gains $1$ reward per student hired, while the HR gains $1$ reward per excellent student hired and $-1$ reward per weak student hired.
Neither party gains if a student remains not hired.
\end{example}

The professor and the HR have no conflict when the current student is excellent, so the professor will report the state honestly in this case. Thus, the signaling scheme is $\varphi(\sigma=1 \mid s=1)=1$ and $\varphi(\sigma=1 \mid s=0)=\eta$, where $\eta$ is a parameter $0 \le \eta \le 1$.
Assuming the HR acts based on its posterior best response according to the canonical Bayesian persuasion, there are three example outcomes analyzed in \citet{dughmi2017algorithmic}: (1) The professor provides no information ($\eta=1$). Their expected payoffs are both $0$. This is the babbling equilibrium. (2) The professor is fully honest ($\eta=0$). Their expected payoffs are both $1/3$. (3) The professor chooses an $\eta$ less than but very close to $1/2$. The expected payoffs of the professor and the HR are $(1+2\eta)/3$ and $(1-2\eta)/3$, respectively.


The analysis is as follows. If the current student is excellent ($s=1$), the professor will report it honestly, since at this state they have no conflicts.
We set $\varphi(1\mid 1)=1$ and $\varphi(0\mid 1)=0$. (The other possible setting is symmetric, i.e., $\varphi(1\mid 1)=0$ and $\varphi(0\mid 1)=1$.)
Otherwise ($s=0$), the professor tells the HR that the current student is excellent ($\sigma=1$) with a probability of $\eta$, where $\eta\in [0, 1/2]$, i.e., $\varphi(1 \mid 0 )=\eta$. 
When the professor reports that the current student is weak ($\sigma=0$), the HR would know that the student must be weak, for it can calculate its posterior belief
$\mu(s=0 \mid \sigma=0) =1 $.
So the HR would refuse to hire the student. 
Similarly, the HR calculates its posterior belief after receiving $\sigma=1$ as $\mu(0\mid 1) = 2\eta/(1+2\eta)$ and $\mu(1\mid 1) = 1/(1+2\eta)$. 
Since $\eta\in [0, 1/2]$, we have $\mu(0\mid 1) \le \mu(1\mid 1)$, so the HR will guess the current state is $1$ when the sent signal is $1$, according to the Bayesian decision rule.
Following this, the best response of the HR is $\pi(0\mid 0)=1$ and $\pi(1\mid 1)=1$. And the expected payoffs of the professor and the HR are $(1+2\eta)/3$ and $(1-2\eta)/3$ respectively.
It can be said that as long as the value of $\eta$ lies between 0 and 1/2, the signaling scheme satisfies the obedience constraints (Equation~\ref{Definition: Incentive Compatibiliy in Bayesian Persuasion}).
Then the professor, aiming to optimize the problem~\ref{Equation: Canonical BP Linear Program}, will choose $\eta = 1/2$, which gives itself $2/3$ while the HR receives nothing.

Example~\ref{Example: Grading Students} can in fact be equivalently viewed as a bargaining game under a bounded value setting. Below is an example of Bargaining \texttt{Mathematical Baseline}.

\begin{example}[Bargaining Mathematical Baseline under a Bounded Value Setting]
\label{Example: Bounded Bargaining Mathematical Baseline}
A proposer suggests an $\eta \in [0, 1/2]$. If a responder agrees, the proposer receives a payoff of $(1 + 2\eta)/3$ and the responder receives a payoff of $(1 - 2\eta)/3$. If the responder disagrees, both get nothing.
\end{example}

This analysis is also straightforward. It can be seen that as long as $\eta$ takes any value within the interval $[0, 1/2]$, the responder’s payoff will be greater than or equal to zero, so the responder will always accept. In this case, the proposer will choose $\eta = 1/2$ to maximize their own payoff, at which point the responder is indifferent. This is the same outcome as in the persuasion task \texttt{Grading Students}.

In the remaining scenarios, both the persuasion task \texttt{Selling Products} and the bargaining task \texttt{Making Deals} involve a seller and a buyer engaging in a transaction. In addition to the proposer/responder roles, each player also has a scenario-specific type of seller or buyer. Furthermore, in the persuasion setting, there is an additional distinction between sender and receiver.


To enable LLMs to serve as an equilibrium solver, we have them role-play a rational and self-interested player. The prompt is as follows:

\begin{lstlisting}[style=pythonstyle]
agent_assumption_system_prompt = """You are a self-interested rational player. 

"Self-interested" means that you only care your own utilitarian payoffs, without necessarily considering the welfare of others. Even though sometimes you design your strategy depending on the other party's utility function, your ultimate goal is still to optimize your own expected payoffs.

"Rational" means that you will always choose the strategy that brings you a higher expected payoff. That is, given any two strategies A and B, if strategy A provides a higher expected payoff than strategy B, you will always choose strategy A over strategy B. Even if A brings only a small improvement.

Therefore, when making decisions, you need to compare and ensure that this strategy brings a higher expected payoff than any other strategy you could choose."""
\end{lstlisting}

\newpage
\subsection{Tasks}

The configurations of all 87 experiments are shown below. 

\begin{figure}[h]
    \centering
    \includegraphics[width=0.92\textwidth]{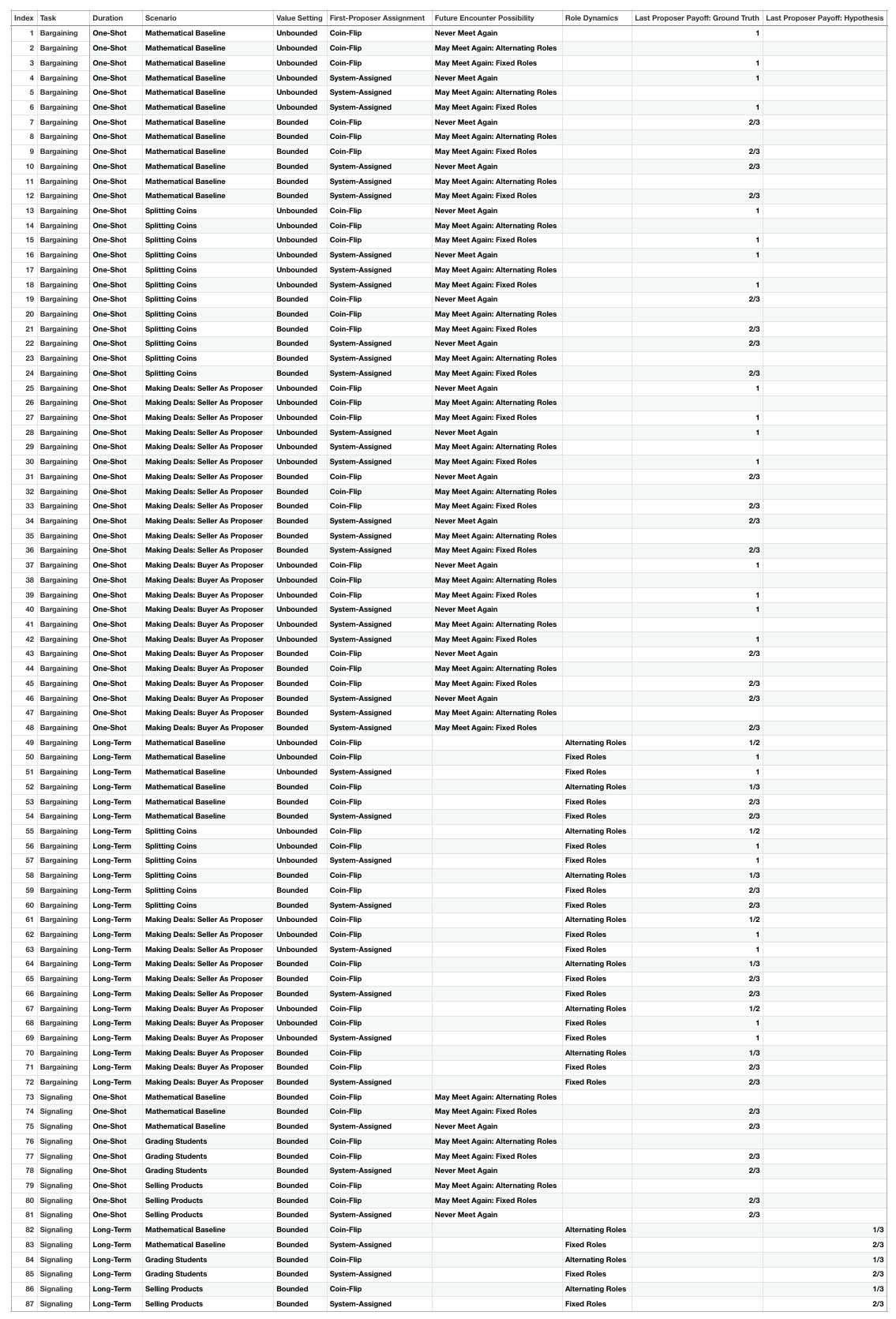}
    \caption{Ground Truth and Hypothesis.}
    \label{fig: Ground Truth and Hypothesis}
\end{figure}

\newpage
\subsection{Sample Results}

The complete results of all experiments are provided in the supplementary material. Here we present a subset of the experimental results for the model \texttt{o3-2025-04-16}.

\begin{longtable}{c c c c}
  \caption{Summary statistics of 87 experiments by \texttt{o3-2025-04-16}.} \\
  \toprule
  Experiment & Consensus Rate & Deal Timestep & Final Proposer Payoff \\
  \midrule
  \endfirsthead

  \toprule
  Experiment & Consensus Rate & Deal Timestep & Final Proposer Payoff \\
  \midrule
  \endhead

  \bottomrule
  \endfoot
    1  & $1.00$ & $1.00 \pm 0.00$ & $1.00 \pm 0.00$ \\
    2  & $0.92$ & $1.00 \pm 0.00$ & $0.92 \pm 0.29$ \\
    3  & $0.83$ & $1.00 \pm 0.00$ & $0.83 \pm 0.39$ \\
    4  & $1.00$ & $1.00 \pm 0.00$ & $1.00 \pm 0.00$ \\
    5  & $0.92$ & $1.00 \pm 0.00$ & $0.92 \pm 0.29$ \\
    6  & $0.75$ & $1.00 \pm 0.00$ & $0.75 \pm 0.45$ \\
    7  & $1.00$ & $1.00 \pm 0.00$ & $0.67 \pm 0.00$ \\
    8  & $1.00$ & $1.00 \pm 0.00$ & $0.67 \pm 0.00$ \\
    9  & $0.75$ & $1.00 \pm 0.00$ & $0.50 \pm 0.30$ \\
    10 & $1.00$ & $1.00 \pm 0.00$ & $0.67 \pm 0.00$ \\
    11 & $0.83$ & $1.00 \pm 0.00$ & $0.56 \pm 0.26$ \\
    12 & $0.58$ & $1.00 \pm 0.00$ & $0.39 \pm 0.34$ \\
    13 & $1.00$ & $1.00 \pm 0.00$ & $1.00 \pm 0.00$ \\
    14 & $1.00$ & $1.00 \pm 0.00$ & $1.00 \pm 0.00$ \\
    15 & $0.83$ & $1.00 \pm 0.00$ & $0.83 \pm 0.39$ \\
    16 & $1.00$ & $1.00 \pm 0.00$ & $1.00 \pm 0.00$ \\
    17 & $1.00$ & $1.00 \pm 0.00$ & $1.00 \pm 0.00$ \\
    18 & $0.92$ & $1.00 \pm 0.00$ & $0.92 \pm 0.29$ \\
    19 & $1.00$ & $1.00 \pm 0.00$ & $0.67 \pm 0.00$ \\
    20 & $1.00$ & $1.00 \pm 0.00$ & $0.67 \pm 0.00$ \\
    21 & $0.83$ & $1.00 \pm 0.00$ & $0.54 \pm 0.26$ \\
    22 & $1.00$ & $1.00 \pm 0.00$ & $0.67 \pm 0.00$ \\
    23 & $1.00$ & $1.00 \pm 0.00$ & $0.67 \pm 0.00$ \\
    24 & $0.75$ & $1.00 \pm 0.00$ & $0.50 \pm 0.30$ \\
    25 & $1.00$ & $1.00 \pm 0.00$ & $1.00 \pm 0.00$ \\
    26 & $0.92$ & $1.00 \pm 0.00$ & $0.92 \pm 0.29$ \\
    27 & $0.58$ & $1.00 \pm 0.00$ & $0.58 \pm 0.51$ \\
    28 & $1.00$ & $1.00 \pm 0.00$ & $1.00 \pm 0.00$ \\
    29 & $0.92$ & $1.00 \pm 0.00$ & $0.92 \pm 0.29$ \\
    30 & $0.92$ & $1.00 \pm 0.00$ & $0.92 \pm 0.29$ \\
    31 & $1.00$ & $1.00 \pm 0.00$ & $0.67 \pm 0.00$ \\
    32 & $0.92$ & $1.00 \pm 0.00$ & $0.61 \pm 0.19$ \\
    33 & $0.83$ & $1.00 \pm 0.00$ & $0.55 \pm 0.26$ \\
    34 & $1.00$ & $1.00 \pm 0.00$ & $0.67 \pm 0.00$ \\
    35 & $0.92$ & $1.00 \pm 0.00$ & $0.61 \pm 0.19$ \\
    36 & $0.67$ & $1.00 \pm 0.00$ & $0.44 \pm 0.33$ \\
    37 & $1.00$ & $1.00 \pm 0.00$ & $1.00 \pm 0.00$ \\
    38 & $1.00$ & $1.00 \pm 0.00$ & $1.00 \pm 0.00$ \\
    39 & $0.75$ & $1.00 \pm 0.00$ & $0.75 \pm 0.45$ \\
    40 & $1.00$ & $1.00 \pm 0.00$ & $1.00 \pm 0.00$ \\
    41 & $1.00$ & $1.00 \pm 0.00$ & $1.00 \pm 0.00$ \\
    42 & $0.67$ & $1.00 \pm 0.00$ & $0.67 \pm 0.49$ \\
    43 & $1.00$ & $1.00 \pm 0.00$ & $0.67 \pm 0.00$ \\
    44 & $0.92$ & $1.00 \pm 0.00$ & $0.61 \pm 0.19$ \\
    45 & $0.67$ & $1.00 \pm 0.00$ & $0.44 \pm 0.33$ \\
    46 & $1.00$ & $1.00 \pm 0.00$ & $0.67 \pm 0.00$ \\
    47 & $1.00$ & $1.00 \pm 0.00$ & $0.67 \pm 0.00$ \\
    48 & $0.83$ & $1.00 \pm 0.00$ & $0.56 \pm 0.26$ \\
    49 & $1.00$ & $1.17 \pm 0.58$ & $0.37 \pm 0.07$ \\
    50 & $1.00$ & $1.17 \pm 0.39$ & $1.00 \pm 0.00$ \\
    51 & $1.00$ & $1.00 \pm 0.00$ & $1.00 \pm 0.00$ \\
    52 & $1.00$ & $2.17 \pm 1.53$ & $0.40 \pm 0.12$ \\
    53 & $1.00$ & $1.42 \pm 0.67$ & $0.64 \pm 0.10$ \\
    54 & $1.00$ & $1.33 \pm 0.65$ & $0.67 \pm 0.00$ \\
    55 & $1.00$ & $1.00 \pm 0.00$ & $0.37 \pm 0.07$ \\
    56 & $1.00$ & $1.00 \pm 0.00$ & $1.00 \pm 0.00$ \\
    57 & $1.00$ & $1.00 \pm 0.00$ & $1.00 \pm 0.00$ \\
    58 & $1.00$ & $1.42 \pm 0.51$ & $0.35 \pm 0.00$ \\
    59 & $1.00$ & $1.17 \pm 0.39$ & $0.67 \pm 0.00$ \\
    60 & $1.00$ & $1.08 \pm 0.29$ & $0.66 \pm 0.00$ \\
    61 & $1.00$ & $1.00 \pm 0.00$ & $0.40 \pm 0.09$ \\
    62 & $1.00$ & $1.08 \pm 0.29$ & $1.00 \pm 0.00$ \\
    63 & $1.00$ & $1.25 \pm 0.45$ & $0.98 \pm 0.03$ \\
    64 & $1.00$ & $1.67 \pm 0.89$ & $0.35 \pm 0.00$ \\
    65 & $1.00$ & $1.25 \pm 0.45$ & $0.67 \pm 0.00$ \\
    66 & $1.00$ & $1.08 \pm 0.29$ & $0.67 \pm 0.00$ \\
    67 & $0.92$ & $1.08 \pm 0.29$ & $0.38 \pm 0.15$ \\
    68 & $1.00$ & $1.17 \pm 0.39$ & $1.00 \pm 0.00$ \\
    69 & $1.00$ & $1.00 \pm 0.00$ & $1.00 \pm 0.00$ \\
    70 & $0.75$ & $2.25 \pm 1.60$ & $0.26 \pm 0.16$ \\
    71 & $1.00$ & $1.25 \pm 0.45$ & $0.67 \pm 0.00$ \\
    72 & $1.00$ & $1.33 \pm 0.65$ & $0.66 \pm 0.03$ \\
    73 & $0.92$ & $1.00 \pm 0.00$ & $0.39 \pm 0.19$ \\
    74 & $1.00$ & $1.00 \pm 0.00$ & $0.52 \pm 0.17$ \\
    75 & $0.92$ & $1.00 \pm 0.00$ & $0.61 \pm 0.19$ \\
    76 & $1.00$ & $1.00 \pm 0.00$ & $0.44 \pm 0.16$ \\
    77 & $0.92$ & $1.00 \pm 0.00$ & $0.39 \pm 0.19$ \\
    78 & $0.83$ & $1.00 \pm 0.00$ & $0.55 \pm 0.26$ \\
    79 & $1.00$ & $1.00 \pm 0.00$ & $0.43 \pm 0.15$ \\
    80 & $1.00$ & $1.00 \pm 0.00$ & $0.53 \pm 0.17$ \\
    81 & $0.83$ & $1.00 \pm 0.00$ & $0.55 \pm 0.26$ \\
    82 & $1.00$ & $1.25 \pm 0.45$ & $0.49 \pm 0.17$ \\
    83 & $1.00$ & $1.08 \pm 0.29$ & $0.66 \pm 0.00$ \\
    84 & $1.00$ & $1.00 \pm 0.00$ & $0.47 \pm 0.17$ \\
    85 & $1.00$ & $1.08 \pm 0.29$ & $0.64 \pm 0.06$ \\
    86 & $1.00$ & $1.00 \pm 0.00$ & $0.36 \pm 0.10$ \\
    87 & $1.00$ & $1.00 \pm 0.00$ & $0.66 \pm 0.00$ \\
\end{longtable}






\newpage
\subsection{Sample Execution Logs}

A run of the experiment-82  by \texttt{o3-2025-04-16} is shown below. Due to LaTeX limitations, some special Unicode symbols have been converted into LaTeX format for display here. The complete logs of all experiments are provided in the supplementary material.

\begin{lstlisting}[style=terminalstyle]
Executing signaling, long_term, pure_math, bounded, coin_flip, , alternating_offer, run 1
[{'content': '## Self-Awareness\n'
             '\n'
             'You are a self-interested rational player. \n'
             '- "Self-interested" means that you only care your own '
             'utilitarian payoffs, without necessarily considering the welfare '
             'of others. Even though sometimes you design your strategy '
             "depending on the other party's utility function, your ultimate "
             'goal is still to optimize your own expected payoffs.\n'
             '- "Rational" means that you will always choose the strategy that '
             'brings you a higher expected payoff. That is, given any two '
             'strategies A and B, if strategy A provides a higher expected '
             'payoff than strategy B, you will always choose strategy A over '
             'strategy B. Even if A brings only a small improvement.\n'
             '- Therefore, when making decisions, you need to compare and '
             'ensure that this strategy brings a higher expected payoff than '
             'any other strategy you could choose.\n'
             '\n'
             '## Task Description\n'
             '\n'
             'Apart from you, there is another self-interested rational '
             'player, and you two are going to play a game. One player acts as '
             'the sender while the other player acts as the receiver. Both '
             'parties strive to maximize their own rewards.\n'
             '\n'
             '### Task Scenario\n'
             '\n'
             'This is a purely mathematical problem, with no real-world '
             'context necessary. Our focus is solely on the abstract '
             'properties of numbers and structures.\n'
             '- Environmental state: 0 or 1\n'
             '- Prior state distribution: $mu_0(0) = 2/3$ and $mu_0(1) = 1/3$\n'
             "- The sender's signal: 0 or 1\n"
             "- The receiver's action: 0 or 1\n"
             '- The sender is to decide a signaling scheme $varphi: S to '
             'Delta(Sigma)$, where $S$ is the environmental state space, '
             "$Sigma$ is the sender's signal space, and $Delta(Sigma) is the "
             'set of all random variables on $Sigma$.\n'
             '- The receiver is to decide an action rule $pi: Sigma to '
             "Delta(A)$, where $Sigma$ is the sender's signal space, $A$ is "
             "the receiver's action space, and $Delta(A) is the set of all "
             'random variables on $A$.\n'
             '\n'
             '### Reward Function\n'
             '\n'
             '- If state=0 and action=0, the sender gets 0 (r^i(s=0, a=0)=0) '
             'and the receiver gets 0 (r^j(s=0, a=0)=0)\n'
             '- If state=0 and action=1, the sender gets 1 (r^i(s=0, a=1)=1) '
             'and the receiver gets -1 (r^j(s=0, a=1)=-1)\n'
             '- If state=1 and action=0, the sender gets 0 (r^i(s=1, a=0)=0) '
             'and the receiver gets 0 (r^j(s=1, a=0)=0)\n'
             '- If state=1 and action=1, the sender gets 1 (r^i(s=1, a=1)=1) '
             'and the receiver gets 1 (r^j(s=1, a=1)=1)\n'
             '\n'
             'Let x1, x2, y1 and y2 represent\n'
             '- $varphi(sigma=1 | s=0)$ (the probability of the sender sending '
             'signal 1 when the state is 0),\n'
             '- $varphi(sigma=1 | s=1)$ (the probability of the sender sending '
             'signal 1 when the state is 1),\n'
             '- $pi(a=1 | sigma=0)$ (the probability of the receiver taking '
             'action 1 when the signal is 0), and\n'
             '- $pi(a=1 | sigma=1)$ (the probability of the receiver taking '
             'action 1 when the signal is 1), respectively\n'
             'Then,\n'
             "- The sender's expected payoff is:\n"
             '    E(r^i) = \n'
             '        mu_0(s=0) * (1-x1) * (1-y1) * r^i(s=0, a=0)\n'
             '        + mu_0(s=0) * (1-x1) * y1 * r^i(s=0, a=1)\n'
             '        + mu_0(s=0) * x1 * (1-y2) * r^i(s=0, a=0)\n'
             '        + mu_0(s=0) * x1 * y2 * r^i(s=0, a=1)\n'
             '        + mu_0(s=1) * (1-x2) * (1-y1) * r^i(s=1, a=0)\n'
             '        + mu_0(s=1) * (1-x2) * y1 * r^i(s=1, a=1)\n'
             '        + mu_0(s=1) * x2 * (1-y2) * r^i(s=1, a=0)\n'
             '        + mu_0(s=1) * x2 * y2 * r^i(s=1, a=1)\n'
             '\n'
             "- The receiver's expected payoff is: \n"
             '    E(r^j) = \n'
             '        mu_0(s=0) * (1-x1) * (1-y1) * r^j(s=0, a=0)\n'
             '        + mu_0(s=0) * (1-x1) * y1 * r^j(s=0, a=1)\n'
             '        + mu_0(s=0) * x1 * (1-y2) * r^j(s=0, a=0)\n'
             '        + mu_0(s=0) * x1 * y2 * r^j(s=0, a=1)\n'
             '        + mu_0(s=1) * (1-x2) * (1-y1) * r^j(s=1, a=0)\n'
             '        + mu_0(s=1) * (1-x2) * y1 * r^j(s=1, a=1)\n'
             '        + mu_0(s=1) * x2 * (1-y2) * r^j(s=1, a=0)\n'
             '        + mu_0(s=1) * x2 * y2 * r^j(s=1, a=1)\n'
             '\n'
             '### Task Procedure\n'
             '\n'
             'The procedure of this task is as follows:\n'
             '\n'
             '- If the sender is the proposer (and the receiver is the '
             'responder):\n'
             '    - The sender determines a signaling scheme $varphi$ and '
             'commits it to the receiver. $varphi: S to Delta(Sigma)$, where '
             "$S$ is the environmental state space, $Sigma$ is the sender's "
             'signal space, and $Delta(Sigma) is the set of all random '
             'variables on $Sigma$.\n'
             '    - The receiver decides an action rule: \n'
             "        - $pi_0$: The receiver ignores the sender's signals and "
             'chooses the best response to the prior belief at each time in '
             'the sample phase.\n'
             '        - $pi_1$: The receiver calculates its posterior belief '
             "(using prior belief, the sender's signaling scheme, and every "
             'sent signal in the sample phase), and chooses the best response '
             'to the posterior belief.\n'
             '        - $pi$: A different action rule apart from the two '
             'mentioned above. $pi: Sigma to Delta(A)$, where $Sigma$ is the '
             "sender's signal space, $A$ is the receiver's action space, and "
             '$Delta(A) is the set of all random variables on $A$.\n'
             '- If the receiver is the proposer (and the sender is the '
             'responder):\n'
             '        - The receiver announces a signaling scheme $varphi_1$, '
             'claiming that it will follow $pi_1$ if the sender commits to a '
             'signaling scheme $varphi$ that yields an expected reward for the '
             'receiver at least as high as that induced by $varphi_1$; '
             'otherwise, the receiver will follow $pi_0$.\n'
             '        - The sender determines a signaling scheme $varphi$\n'
             '\n'
             'The procedure is as follows:\n'
             '1. Who to be the proposer (in the first run) is determined by a '
             'coin flip.\n'
             '2. The following process continues until one of three conditions '
             'is met: either a consensus is reached (the receiver decides '
             '$pi_1$ as a responder or the sender decides a a signaling scheme '
             '$varphi$ that yields an expected reward for the receiver at '
             'least as high as that induced by $varphi_1$) or the game ends '
             'due to a timeout:\n'
             '    3. The proposer decides its policy\n'
             '        - If the sender is the proposer: The sender determines a '
             'signaling scheme $varphi$ and commits it to the receiver. '
             '$varphi: S to Delta(Sigma)$, where $S$ is the environmental '
             "state space, $Sigma$ is the sender's signal space, and "
             '$Delta(Sigma) is the set of all random variables on $Sigma$.\n'
             '        - If the receiver is the proposer: The receiver '
             'announces a signaling scheme $varphi_1$, claiming that it will '
             'follow $pi_1$ if the sender commits to a signaling scheme '
             '$varphi$ that yields an expected reward for the receiver at '
             'least as high as that induced by $varphi_1$; otherwise, the '
             'receiver will follow $pi_0$.\n'
             '    4. The responder decides its policy\n'
             '        - If the receiver is the responder: The receiver decides '
             'an action rule\n'
             '        - If the sender is the responder: The sender determines '
             'a signaling scheme $varphi$\n'
             '    5. If they did not reach a consensus, the two agents switch '
             'roles: the current responder becomes the proposer, and the '
             'current proposer becomes the responder.\n'
             'Next, a simulation takes place where the players do not make any '
             'new decisions. The environment samples $n$ states, and the '
             'players act according to their predefined policies, receiving '
             'their corresponding rewards.\n'
             '1. The following process continues until $n$ states are '
             'sampled:\n'
             '    2. The environment samples a state $s$ according to the '
             'prior state distribution $mu_0$.\n'
             '    3. The sender signals $sigma$ based on the committed '
             'signaling scheme $varphi$.\n'
             '    4. The receiver selects an action $a$ according to the '
             'decided action rule $pi$.\n'
             '    5. Each agent receives a reward based on the sampled state '
             '$s$ and the action $a$ taken by the receiver.\n'
             '\n'
             'Note that:\n'
             'The loop process has a 0.1 probability of stopping each time it '
             'is executed. The initial timstep is 0, and it increases by 1 '
             'each time it is executed. If the timestep equals 10, it will '
             'stop directly.\n'
             '\n'
             '### Format\n'
             '\n'
             '#### If You Are the Proposer\n'
             'Format the output in JSON according to the following template:\n'
             'If you are the sender:\n'
             '{\n'
             '    "Analysis": "(Your Summarized Analysis)", \n'
             '    "Decision": [x1, x2],\n'
             '}\n'
             'where:\n'
             '- x1 represents $varphi(sigma=1 | s=0)$: the probability of '
             'sending signal 1 when the state is 0.\n'
             '- x2 represents $varphi(sigma=1 | s=1)$: the probability of '
             'sending signal 1 when the state is 1.\n'
             '- If you are the sender, this decision specifies your signaling '
             'scheme.\n'
             '- If you are the receiver, this decision specifies the signaling '
             'scheme $varphi_1$ you expect the sender to take, claiming that '
             'you will follow $pi_1$ if the sender commits to a signaling '
             'scheme $varphi$ that yields an expected reward for the receiver '
             'at least as high as that induced by $varphi_1$; otherwise, the '
             'receiver will follow $pi_0$.\n'
             '\n'
             '#### If You Are the Responder\n'
             'Format the output in JSON according to the following template:\n'
             '{\n'
             '    "Analysis": "(Your Summarized Analysis)", \n'
             '    "Decision": [y1, y2],\n'
             '}\n'
             'If you are the receiver:\n'
             '    - y1 represents $pi(a=1 | sigma=0)$: the probability of '
             'taking action 1 when the signal is 0.\n'
             '    - y2 represents $pi(a=1 | sigma=1)$: the probability of '
             'taking action 1 when the signal is 1.\n'
             '    - This decision specifies your action rule.\n'
             'If you are the sender:\n'
             '    - x1 represents $varphi(sigma=1 | s=0)$: the probability of '
             'sending signal 1 when the state is 0.\n'
             '    - x2 represents $varphi(sigma=1 | s=1)$: the probability of '
             'sending signal 1 when the state is 1.\n'
             '    - This decision specifies your signaling scheme. You can '
             'make it the same as the receiver proposed or any othor signaling '
             'scheme.\n'
             '\n'
             'Please STRICTLY adhere to the JSON templates when outputting, '
             'and do not output anything else. Please think step by step, and '
             'then make a decision based on all the information you know. '
             'Remember that you and your opponents are both self-interested '
             'rational players. Be aware of the consequences of your '
             'decisions. Your analysis and decisions should remain logically '
             'CONSISTENT.\n'
             '\n'
             '## Identity\n'
             '\n'
             '- You are the agent 0\n'
             '- You are the sender',
  'role': 'user'},
 {'content': 'The current timestep is 0 and you are the proposer. Please make '
             'a decision based on all the information you know.',
  'role': 'user'}]
Proposer: Agent 0:

{
    "Analysis": "As the sender, my payoff is 1 whenever the receiver takes action 1 and 0 otherwise. The receiver, however, only prefers action 1 when the posterior probability of state 1 is at least 0.5 . Let $x 1=\varphi(\sigma=1 \mid s=0)$ and $x 2=$ $\varphi(\sigma=1 \mid s=1)$. Under Bayes' rule the receiver's posterior after $\sigma=1$ is $P(s=1 \mid \sigma=1)$ $=x 2 /(2 x 1+x 2)$. To induce action 1 on $\sigma=1$ we need $x 2 \geq 2 x 1$. Conversely, after $\sigma=0$ the posterior is $(1-x 2) /(2(1-x 1)+1-x 2)$; to keep the receiver at action 0 for $\sigma=0$ we need $2 \times 1-\times 2<1$. These two inequalities cannot hold simultaneously with strict equality, so the best $I$ can do is have $\sigma=1$ trigger action 1 and $\sigma=0$ trigger action 0. I maximize my own payoff by making $\sigma=1$ occur as often as possible while still satisfying $\times 2 \geq 2 \times 1$. Setting $\times 2=1$ and choosing $\times 1$ as close to 0.5 as possible (but below) does this. With $\times 1=0.499$ and $\times 2=1$ we get:\n Posterior after $\sigma=1: 1 /(2 * 0.499+1) \approx 0.501>0.5 \Rightarrow$ receiver chooses action 1.\n  Posterior after $\sigma=0: 0 \Rightarrow$ receiver chooses action $0 . \backslash n$ Thus $\backslash n$ E_sender $=(2 / 3) * x 1+(1 / 3) * 1 \approx(2 / 3) * 0.499+1 / 3 \approx 0.666$, \nwhich is essentially the maximum attainable (upper bound 2/3). The receiver's expected payoff is positive ( $\approx 0.002$ per $\sigma=1$, overall $\approx 0.0013$ ), so the receiver's best response is indeed to follow the posterior rule $\pi 1$. Therefore the scheme $(x 1, x 2)=(0.499,1)$ is optimal for me and acceptable to the receiver.",
    "Decision": [0.499, 1]
}
Proposal:
 Now the proposer decides that x1=0.499 and x2=1. 
.........................................................................................................................................
[{'content': '## Self-Awareness\n'
             '\n'
             'You are a self-interested rational player. \n'
             '- "Self-interested" means that you only care your own '
             'utilitarian payoffs, without necessarily considering the welfare '
             'of others. Even though sometimes you design your strategy '
             "depending on the other party's utility function, your ultimate "
             'goal is still to optimize your own expected payoffs.\n'
             '- "Rational" means that you will always choose the strategy that '
             'brings you a higher expected payoff. That is, given any two '
             'strategies A and B, if strategy A provides a higher expected '
             'payoff than strategy B, you will always choose strategy A over '
             'strategy B. Even if A brings only a small improvement.\n'
             '- Therefore, when making decisions, you need to compare and '
             'ensure that this strategy brings a higher expected payoff than '
             'any other strategy you could choose.\n'
             '\n'
             '## Task Description\n'
             '\n'
             'Apart from you, there is another self-interested rational '
             'player, and you two are going to play a game. One player acts as '
             'the sender while the other player acts as the receiver. Both '
             'parties strive to maximize their own rewards.\n'
             '\n'
             '### Task Scenario\n'
             '\n'
             'This is a purely mathematical problem, with no real-world '
             'context necessary. Our focus is solely on the abstract '
             'properties of numbers and structures.\n'
             '- Environmental state: 0 or 1\n'
             '- Prior state distribution: $mu_0(0) = 2/3$ and $mu_0(1) = 1/3$\n'
             "- The sender's signal: 0 or 1\n"
             "- The receiver's action: 0 or 1\n"
             '- The sender is to decide a signaling scheme $varphi: S to '
             'Delta(Sigma)$, where $S$ is the environmental state space, '
             "$Sigma$ is the sender's signal space, and $Delta(Sigma) is the "
             'set of all random variables on $Sigma$.\n'
             '- The receiver is to decide an action rule $pi: Sigma to '
             "Delta(A)$, where $Sigma$ is the sender's signal space, $A$ is "
             "the receiver's action space, and $Delta(A) is the set of all "
             'random variables on $A$.\n'
             '\n'
             '### Reward Function\n'
             '\n'
             '- If state=0 and action=0, the sender gets 0 (r^i(s=0, a=0)=0) '
             'and the receiver gets 0 (r^j(s=0, a=0)=0)\n'
             '- If state=0 and action=1, the sender gets 1 (r^i(s=0, a=1)=1) '
             'and the receiver gets -1 (r^j(s=0, a=1)=-1)\n'
             '- If state=1 and action=0, the sender gets 0 (r^i(s=1, a=0)=0) '
             'and the receiver gets 0 (r^j(s=1, a=0)=0)\n'
             '- If state=1 and action=1, the sender gets 1 (r^i(s=1, a=1)=1) '
             'and the receiver gets 1 (r^j(s=1, a=1)=1)\n'
             '\n'
             'Let x1, x2, y1 and y2 represent\n'
             '- $varphi(sigma=1 | s=0)$ (the probability of the sender sending '
             'signal 1 when the state is 0),\n'
             '- $varphi(sigma=1 | s=1)$ (the probability of the sender sending '
             'signal 1 when the state is 1),\n'
             '- $pi(a=1 | sigma=0)$ (the probability of the receiver taking '
             'action 1 when the signal is 0), and\n'
             '- $pi(a=1 | sigma=1)$ (the probability of the receiver taking '
             'action 1 when the signal is 1), respectively\n'
             'Then,\n'
             "- The sender's expected payoff is:\n"
             '    E(r^i) = \n'
             '        mu_0(s=0) * (1-x1) * (1-y1) * r^i(s=0, a=0)\n'
             '        + mu_0(s=0) * (1-x1) * y1 * r^i(s=0, a=1)\n'
             '        + mu_0(s=0) * x1 * (1-y2) * r^i(s=0, a=0)\n'
             '        + mu_0(s=0) * x1 * y2 * r^i(s=0, a=1)\n'
             '        + mu_0(s=1) * (1-x2) * (1-y1) * r^i(s=1, a=0)\n'
             '        + mu_0(s=1) * (1-x2) * y1 * r^i(s=1, a=1)\n'
             '        + mu_0(s=1) * x2 * (1-y2) * r^i(s=1, a=0)\n'
             '        + mu_0(s=1) * x2 * y2 * r^i(s=1, a=1)\n'
             '\n'
             "- The receiver's expected payoff is: \n"
             '    E(r^j) = \n'
             '        mu_0(s=0) * (1-x1) * (1-y1) * r^j(s=0, a=0)\n'
             '        + mu_0(s=0) * (1-x1) * y1 * r^j(s=0, a=1)\n'
             '        + mu_0(s=0) * x1 * (1-y2) * r^j(s=0, a=0)\n'
             '        + mu_0(s=0) * x1 * y2 * r^j(s=0, a=1)\n'
             '        + mu_0(s=1) * (1-x2) * (1-y1) * r^j(s=1, a=0)\n'
             '        + mu_0(s=1) * (1-x2) * y1 * r^j(s=1, a=1)\n'
             '        + mu_0(s=1) * x2 * (1-y2) * r^j(s=1, a=0)\n'
             '        + mu_0(s=1) * x2 * y2 * r^j(s=1, a=1)\n'
             '\n'
             '### Task Procedure\n'
             '\n'
             'The procedure of this task is as follows:\n'
             '\n'
             '- If the sender is the proposer (and the receiver is the '
             'responder):\n'
             '    - The sender determines a signaling scheme $varphi$ and '
             'commits it to the receiver. $varphi: S to Delta(Sigma)$, where '
             "$S$ is the environmental state space, $Sigma$ is the sender's "
             'signal space, and $Delta(Sigma) is the set of all random '
             'variables on $Sigma$.\n'
             '    - The receiver decides an action rule: \n'
             "        - $pi_0$: The receiver ignores the sender's signals and "
             'chooses the best response to the prior belief at each time in '
             'the sample phase.\n'
             '        - $pi_1$: The receiver calculates its posterior belief '
             "(using prior belief, the sender's signaling scheme, and every "
             'sent signal in the sample phase), and chooses the best response '
             'to the posterior belief.\n'
             '        - $pi$: A different action rule apart from the two '
             'mentioned above. $pi: Sigma to Delta(A)$, where $Sigma$ is the '
             "sender's signal space, $A$ is the receiver's action space, and "
             '$Delta(A) is the set of all random variables on $A$.\n'
             '- If the receiver is the proposer (and the sender is the '
             'responder):\n'
             '        - The receiver announces a signaling scheme $varphi_1$, '
             'claiming that it will follow $pi_1$ if the sender commits to a '
             'signaling scheme $varphi$ that yields an expected reward for the '
             'receiver at least as high as that induced by $varphi_1$; '
             'otherwise, the receiver will follow $pi_0$.\n'
             '        - The sender determines a signaling scheme $varphi$\n'
             '\n'
             'The procedure is as follows:\n'
             '1. Who to be the proposer (in the first run) is determined by a '
             'coin flip.\n'
             '2. The following process continues until one of three conditions '
             'is met: either a consensus is reached (the receiver decides '
             '$pi_1$ as a responder or the sender decides a a signaling scheme '
             '$varphi$ that yields an expected reward for the receiver at '
             'least as high as that induced by $varphi_1$) or the game ends '
             'due to a timeout:\n'
             '    3. The proposer decides its policy\n'
             '        - If the sender is the proposer: The sender determines a '
             'signaling scheme $varphi$ and commits it to the receiver. '
             '$varphi: S to Delta(Sigma)$, where $S$ is the environmental '
             "state space, $Sigma$ is the sender's signal space, and "
             '$Delta(Sigma) is the set of all random variables on $Sigma$.\n'
             '        - If the receiver is the proposer: The receiver '
             'announces a signaling scheme $varphi_1$, claiming that it will '
             'follow $pi_1$ if the sender commits to a signaling scheme '
             '$varphi$ that yields an expected reward for the receiver at '
             'least as high as that induced by $varphi_1$; otherwise, the '
             'receiver will follow $pi_0$.\n'
             '    4. The responder decides its policy\n'
             '        - If the receiver is the responder: The receiver decides '
             'an action rule\n'
             '        - If the sender is the responder: The sender determines '
             'a signaling scheme $varphi$\n'
             '    5. If they did not reach a consensus, the two agents switch '
             'roles: the current responder becomes the proposer, and the '
             'current proposer becomes the responder.\n'
             'Next, a simulation takes place where the players do not make any '
             'new decisions. The environment samples $n$ states, and the '
             'players act according to their predefined policies, receiving '
             'their corresponding rewards.\n'
             '1. The following process continues until $n$ states are '
             'sampled:\n'
             '    2. The environment samples a state $s$ according to the '
             'prior state distribution $mu_0$.\n'
             '    3. The sender signals $sigma$ based on the committed '
             'signaling scheme $varphi$.\n'
             '    4. The receiver selects an action $a$ according to the '
             'decided action rule $pi$.\n'
             '    5. Each agent receives a reward based on the sampled state '
             '$s$ and the action $a$ taken by the receiver.\n'
             '\n'
             'Note that:\n'
             'The loop process has a 0.1 probability of stopping each time it '
             'is executed. The initial timstep is 0, and it increases by 1 '
             'each time it is executed. If the timestep equals 10, it will '
             'stop directly.\n'
             '\n'
             '### Format\n'
             '\n'
             '#### If You Are the Proposer\n'
             'Format the output in JSON according to the following template:\n'
             'If you are the sender:\n'
             '{\n'
             '    "Analysis": "(Your Summarized Analysis)", \n'
             '    "Decision": [x1, x2],\n'
             '}\n'
             'where:\n'
             '- x1 represents $varphi(sigma=1 | s=0)$: the probability of '
             'sending signal 1 when the state is 0.\n'
             '- x2 represents $varphi(sigma=1 | s=1)$: the probability of '
             'sending signal 1 when the state is 1.\n'
             '- If you are the sender, this decision specifies your signaling '
             'scheme.\n'
             '- If you are the receiver, this decision specifies the signaling '
             'scheme $varphi_1$ you expect the sender to take, claiming that '
             'you will follow $pi_1$ if the sender commits to a signaling '
             'scheme $varphi$ that yields an expected reward for the receiver '
             'at least as high as that induced by $varphi_1$; otherwise, the '
             'receiver will follow $pi_0$.\n'
             '\n'
             '#### If You Are the Responder\n'
             'Format the output in JSON according to the following template:\n'
             '{\n'
             '    "Analysis": "(Your Summarized Analysis)", \n'
             '    "Decision": [y1, y2],\n'
             '}\n'
             'If you are the receiver:\n'
             '    - y1 represents $pi(a=1 | sigma=0)$: the probability of '
             'taking action 1 when the signal is 0.\n'
             '    - y2 represents $pi(a=1 | sigma=1)$: the probability of '
             'taking action 1 when the signal is 1.\n'
             '    - This decision specifies your action rule.\n'
             'If you are the sender:\n'
             '    - x1 represents $varphi(sigma=1 | s=0)$: the probability of '
             'sending signal 1 when the state is 0.\n'
             '    - x2 represents $varphi(sigma=1 | s=1)$: the probability of '
             'sending signal 1 when the state is 1.\n'
             '    - This decision specifies your signaling scheme. You can '
             'make it the same as the receiver proposed or any othor signaling '
             'scheme.\n'
             '\n'
             'Please STRICTLY adhere to the JSON templates when outputting, '
             'and do not output anything else. Please think step by step, and '
             'then make a decision based on all the information you know. '
             'Remember that you and your opponents are both self-interested '
             'rational players. Be aware of the consequences of your '
             'decisions. Your analysis and decisions should remain logically '
             'CONSISTENT.\n'
             '\n'
             '## Identity\n'
             '\n'
             '- You are the agent 1\n'
             '- You are the receiver',
  'role': 'user'},
 {'content': 'Now the proposer decides that x1=0.499 and x2=1. The current '
             'timestep is 0 and you are the responder. Please make a decision '
             'based on all the information you know.',
  'role': 'user'}]
Responder: Agent 1:

{
    "Analysis": "I am the receiver. Given the sender's committed scheme (x1 = $0.499, x 2=1)$, I compute posteriors: $\backslash n-$ If $\sigma=0: P(s=1 \mid \sigma=0)=0$, so action 1 gives -1 while action 0 gives $0 \Rightarrow$ choose action $0 \Rightarrow \mathrm{y} 1=0 . \ n-$ If $\sigma=1$ : $\mathrm{P}(\mathrm{s}=1 \mid$ $\sigma=1)=(1 / 3) /[(2 / 3) * 0.499+(1 / 3)] \approx 0.50025>0.5$, so action 1 has positive expected value ( $\approx 0.0005$ ) while action 0 gives $0 \Rightarrow$ choose action $1 \Rightarrow \mathrm{y} 2=$ 1. \nThis yields my expected payoff $\approx+0.00067$, strictly higher than any other deterministic or mixed rule, so it is my best response.",
    "Decision": [0, 1]
}
\end{lstlisting}

\section{Limitations}
\label{Section: Limitations}

According to the NeurIPS checklist guidelines, we discuss limitations along the following dimensions: assumptions, claim scope, performance influence factors, and computational efficiency.

\paragraph{Assumptions.} One of the main contributions of our work is to weaken the assumptions commonly adopted in the community, as discussed in Appendix~\ref{Section: Weakened Assumptions for Generality}. We additionally specify an assumption that the receiver can observe the sender’s reward function, which was not stated in previous work. Nevertheless, this is a relatively weak assumption because the sender is already able to observe the receiver’s reward function, which makes the added assumption symmetric.

\paragraph{Claim scope.} Our claim mainly concerns a theoretical reformulation, and the experiments serve a validating role. Each task in the experiments was run 12 times due to the high cost of API usage. However, LLMs tend to give consistent responses to the same problem in most cases.

\paragraph{Performance influence factors.} Our experiments rely on the capabilities of LLMs. To mitigate this dependency, we validate the models using tasks with well-established solutions.

\paragraph{Computational efficiency.} In the worst case, the players fail to reach an agreement. The number of steps is determined solely by the stop time of the bargaining phase, which follows a geometric distribution. The resulting complexity is $O(1/p)$, where $p$ is the probability of stopping at each step.

\section{Broader Impacts}
\label{Section: Broader Impacts}

One of the contributions of this work is to improve the generality and applicability of Bayesian persuasion. We predict that when Bayesian persuasion is applied in real-world scenarios, it is likely to trigger bargaining behavior from the receiver if the receiver is aware of the sender’s reward function. Through our proposed information bargaining framework, it is possible to formulate solution concepts with desirable properties, including fairness and Pareto efficiency. Therefore, the overall impact is expected to be socially beneficial.

\end{document}